\documentclass[a4paper,11pt]{article}
\pdfoutput=1 

\usepackage{jheppub} 

\usepackage[T1]{fontenc} 
\usepackage{braket}
\usepackage{csquotes}

\newcounter{AffiliationCounter}
\stepcounter{AffiliationCounter}\edef\instBilbao{\protect\theAffiliationCounter}
\stepcounter{AffiliationCounter}\edef\instBNL{\protect\theAffiliationCounter}
\stepcounter{AffiliationCounter}\edef\instBINP{\protect\theAffiliationCounter}
\stepcounter{AffiliationCounter}\edef\instCharles{\protect\theAffiliationCounter}
\stepcounter{AffiliationCounter}\edef\instCincinnati{\protect\theAffiliationCounter}
\stepcounter{AffiliationCounter}\edef\instDESY{\protect\theAffiliationCounter}
\stepcounter{AffiliationCounter}\edef\instDuke{\protect\theAffiliationCounter}
\stepcounter{AffiliationCounter}\edef\instFudan{\protect\theAffiliationCounter}
\stepcounter{AffiliationCounter}\edef\instGoettingen{\protect\theAffiliationCounter}
\stepcounter{AffiliationCounter}\edef\instSokendai{\protect\theAffiliationCounter}
\stepcounter{AffiliationCounter}\edef\instHanyang{\protect\theAffiliationCounter}
\stepcounter{AffiliationCounter}\edef\instHawaii{\protect\theAffiliationCounter}
\stepcounter{AffiliationCounter}\edef\instKEK{\protect\theAffiliationCounter}
\stepcounter{AffiliationCounter}\edef\instJPARC{\protect\theAffiliationCounter}
\stepcounter{AffiliationCounter}\edef\instJuelich{\protect\theAffiliationCounter}
\stepcounter{AffiliationCounter}\edef\instIKER{\protect\theAffiliationCounter}
\stepcounter{AffiliationCounter}\edef\instIISERM{\protect\theAffiliationCounter}
\stepcounter{AffiliationCounter}\edef\instIITB{\protect\theAffiliationCounter}
\stepcounter{AffiliationCounter}\edef\instIITG{\protect\theAffiliationCounter}
\stepcounter{AffiliationCounter}\edef\instIITH{\protect\theAffiliationCounter}
\stepcounter{AffiliationCounter}\edef\instIITM{\protect\theAffiliationCounter}
\stepcounter{AffiliationCounter}\edef\instIndiana{\protect\theAffiliationCounter}
\stepcounter{AffiliationCounter}\edef\instVienna{\protect\theAffiliationCounter}
\stepcounter{AffiliationCounter}\edef\instNapoli{\protect\theAffiliationCounter}
\stepcounter{AffiliationCounter}\edef\instTorino{\protect\theAffiliationCounter}
\stepcounter{AffiliationCounter}\edef\instJAEA{\protect\theAffiliationCounter}
\stepcounter{AffiliationCounter}\edef\instJSI{\protect\theAffiliationCounter}
\stepcounter{AffiliationCounter}\edef\instKarlsruhe{\protect\theAffiliationCounter}
\stepcounter{AffiliationCounter}\edef\instKennesaw{\protect\theAffiliationCounter}
\stepcounter{AffiliationCounter}\edef\instKACST{\protect\theAffiliationCounter}
\stepcounter{AffiliationCounter}\edef\instKAU{\protect\theAffiliationCounter}
\stepcounter{AffiliationCounter}\edef\instKitasato{\protect\theAffiliationCounter}
\stepcounter{AffiliationCounter}\edef\instKISTI{\protect\theAffiliationCounter}
\stepcounter{AffiliationCounter}\edef\instKorea{\protect\theAffiliationCounter}
\stepcounter{AffiliationCounter}\edef\instKyoto{\protect\theAffiliationCounter}
\stepcounter{AffiliationCounter}\edef\instKyungpook{\protect\theAffiliationCounter}
\stepcounter{AffiliationCounter}\edef\instLAL{\protect\theAffiliationCounter}
\stepcounter{AffiliationCounter}\edef\instLausanne{\protect\theAffiliationCounter}
\stepcounter{AffiliationCounter}\edef\instLebedev{\protect\theAffiliationCounter}
\stepcounter{AffiliationCounter}\edef\instLjubljana{\protect\theAffiliationCounter}
\stepcounter{AffiliationCounter}\edef\instLMU{\protect\theAffiliationCounter}
\stepcounter{AffiliationCounter}\edef\instLuther{\protect\theAffiliationCounter}
\stepcounter{AffiliationCounter}\edef\instMaribor{\protect\theAffiliationCounter}
\stepcounter{AffiliationCounter}\edef\instMPI{\protect\theAffiliationCounter}
\stepcounter{AffiliationCounter}\edef\instMelbourne{\protect\theAffiliationCounter}
\stepcounter{AffiliationCounter}\edef\instMississippi{\protect\theAffiliationCounter}
\stepcounter{AffiliationCounter}\edef\instMiyazaki{\protect\theAffiliationCounter}
\stepcounter{AffiliationCounter}\edef\instMEPhI{\protect\theAffiliationCounter}
\stepcounter{AffiliationCounter}\edef\instMIPT{\protect\theAffiliationCounter}
\stepcounter{AffiliationCounter}\edef\instNagoya{\protect\theAffiliationCounter}
\stepcounter{AffiliationCounter}\edef\instNagoyaKMI{\protect\theAffiliationCounter}
\stepcounter{AffiliationCounter}\edef\instUNapoli{\protect\theAffiliationCounter}
\stepcounter{AffiliationCounter}\edef\instNara{\protect\theAffiliationCounter}
\stepcounter{AffiliationCounter}\edef\instNCU{\protect\theAffiliationCounter}
\stepcounter{AffiliationCounter}\edef\instNUU{\protect\theAffiliationCounter}
\stepcounter{AffiliationCounter}\edef\instTaiwan{\protect\theAffiliationCounter}
\stepcounter{AffiliationCounter}\edef\instKrakow{\protect\theAffiliationCounter}
\stepcounter{AffiliationCounter}\edef\instNihonDental{\protect\theAffiliationCounter}
\stepcounter{AffiliationCounter}\edef\instNiigata{\protect\theAffiliationCounter}
\stepcounter{AffiliationCounter}\edef\instNovosibirsk{\protect\theAffiliationCounter}
\stepcounter{AffiliationCounter}\edef\instOsakaCity{\protect\theAffiliationCounter}
\stepcounter{AffiliationCounter}\edef\instPNNL{\protect\theAffiliationCounter}
\stepcounter{AffiliationCounter}\edef\instPanjab{\protect\theAffiliationCounter}
\stepcounter{AffiliationCounter}\edef\instPeking{\protect\theAffiliationCounter}
\stepcounter{AffiliationCounter}\edef\instPittsburgh{\protect\theAffiliationCounter}
\stepcounter{AffiliationCounter}\edef\instPunjab{\protect\theAffiliationCounter}
\stepcounter{AffiliationCounter}\edef\instNPC{\protect\theAffiliationCounter}
\stepcounter{AffiliationCounter}\edef\instRIKEN{\protect\theAffiliationCounter}
\stepcounter{AffiliationCounter}\edef\instUSTC{\protect\theAffiliationCounter}
\stepcounter{AffiliationCounter}\edef\instShoyaku{\protect\theAffiliationCounter}
\stepcounter{AffiliationCounter}\edef\instSoongsil{\protect\theAffiliationCounter}
\stepcounter{AffiliationCounter}\edef\instSungkyunkwan{\protect\theAffiliationCounter}
\stepcounter{AffiliationCounter}\edef\instSydney{\protect\theAffiliationCounter}
\stepcounter{AffiliationCounter}\edef\instTabuk{\protect\theAffiliationCounter}
\stepcounter{AffiliationCounter}\edef\instTata{\protect\theAffiliationCounter}
\stepcounter{AffiliationCounter}\edef\instToho{\protect\theAffiliationCounter}
\stepcounter{AffiliationCounter}\edef\instTohoku{\protect\theAffiliationCounter}
\stepcounter{AffiliationCounter}\edef\instERI{\protect\theAffiliationCounter}
\stepcounter{AffiliationCounter}\edef\instTokyo{\protect\theAffiliationCounter}
\stepcounter{AffiliationCounter}\edef\instTIT{\protect\theAffiliationCounter}
\stepcounter{AffiliationCounter}\edef\instTMU{\protect\theAffiliationCounter}
\stepcounter{AffiliationCounter}\edef\instVPI{\protect\theAffiliationCounter}
\stepcounter{AffiliationCounter}\edef\instWayneState{\protect\theAffiliationCounter}
\stepcounter{AffiliationCounter}\edef\instYamagata{\protect\theAffiliationCounter}
\stepcounter{AffiliationCounter}\edef\instYonsei{\protect\theAffiliationCounter}

\collaboration{The Belle Collaboration}
  \author[\instIITM,\hbox{$\dagger$}]{P.~K.~Resmi,\note[$\dagger$]{Corresponding author.}} 
  \author[\instIITM]{J.~Libby,} 
  \author[\instLAL]{K.~Trabelsi,} 
  \author[\instKEK,\instSokendai]{I.~Adachi,} 
  \author[\instTokyo]{H.~Aihara,} 
  \author[\instTabuk,\instKAU]{S.~Al~Said,} 
  \author[\instBNL]{D.~M.~Asner,} 
  \author[\instBINP,\instNovosibirsk]{V.~Aulchenko,} 
  \author[\instMIPT]{T.~Aushev,} 
 \author[\instDESY]{V.~Babu,} 
% ...

\emailAdd{resmipk@physics.iitm.ac.in}

\title{\boldmath First measurement of the CKM angle $\phi_3$ with $B^{\pm}\to D(K_{\rm S}^0\pi^+\pi^-\pi^0)K^{\pm}$ decays}

\preprint{\vbox{ \hbox{   }
					    	\hbox{Belle Preprint 2019-14}
                        	\hbox{KEK Preprint 2019-19} 
                     }}

\abstract{We present the first model-independent measurement of the CKM unitarity triangle angle $\phi_3$ using $B^{\pm}\to D(K_{\rm S}^0\pi^+\pi^-\pi^0)K^{\pm}$ decays, where $D$ indicates either a $D^{0}$ or $\overline{D}^{0}$ meson.
Measurements of the strong-phase difference of the $D \to K_{\rm S}^0\pi^+\pi^-\pi^0$ amplitude obtained from CLEO-c data are used as input.  This analysis is based on the full Belle data set of $772\times 10^{6}$ $B\overline{B}$ events collected at the $\Upsilon(4S)$ resonance. We obtain  $\phi_3 = (5.7~^{+10.2}_{-8.8} \pm 3.5 \pm 5.7)^{\circ}$ and the suppressed amplitude ratio $r_{B} = 0.323 \pm 0.147 \pm 0.023 \pm 0.051$. Here the first uncertainty is statistical, the second is the experimental systematic, and the third is due to the precision of the strong-phase parameters measured from CLEO-c data. The 95\% confidence interval on $\phi_3$ is $(-29.7,~109.5)^{\circ}$, which is consistent with the current world average.}


\begin{document} 
\maketitle
\flushbottom

\section{Introduction}
\label{sec:intro}
The description of $CP$ violation in the standard model (SM) can be tested via measurements of observables related to the Cabibbo-Kobayashi-Maskawa (CKM) matrix~\cite{C,KM}. One such test is the measurement of the unitarity-triangle angle $\phi_3 \equiv \arg(-V_{ud}V_{ub}^{*}/V_{cd}V_{cb}^{*})$, also denoted as $\gamma$. Here, $V_{ij}$ refers to the CKM matrix element. The angle $\phi_3$ is accessible through tree-level amplitudes, and the associated theoretical uncertainty is negligible $\left[\mathcal{O}(10^{-7})\right]$~\cite{BrodJupan}. A comparison of the direct measurements of $\phi_3$ with the value inferred from other measurements related to the CKM matrix~\cite{CKMfitter}, which are more likely to be influenced by beyond-SM physics~\cite{BSM1,BSM2}, provides a probe for new physics. The current experimental uncertainty on $\phi_3$~\cite{CKMfitter} limits such tests, motivating more precise measurements of the angle. 

The measurement of $\phi_3$ is possible when there is interference between the transitions $\overline{b}\to\overline{c}u\overline{s}$ and $\overline{b}\to \overline{u}c\overline{s}$. This is the case in the decay $B^{+}\to DK^{+}$, where $D$ is a neutral charm meson decaying to a final state common to both $D^{0}$ and $\overline{D}^0$. Here and elsewhere in this paper, inclusion of charge-conjugate final states is implied unless explicitly stated otherwise. Currently, the most precise measurement of $\phi_3$  \cite{LatestLHCb} exploits the self-conjugate final state $D\to K_{\rm S}^0\pi^+\pi^-$, where the $CP$ asymmetry in different regions of the $D$ meson Dalitz plot is measured to determine $\phi_3$~\cite{GGSZ, GGSZ2}. The analysis requires knowledge of the strong-phase difference between the $D^0$ and $\overline{D}^0$ decay amplitudes, and measured values of the strong-phase differences averaged over Dalitz plot bins are used as input~\cite{CLEO-KsPiPi}. Given the success of such analyses in obtaining $\phi_3$~\cite{LatestLHCb, Belle-GGSZ}, other self-conjugate final states can be studied in a similar fashion to improve the determination of $\phi_3$.   

In this paper, we present the first measurements of the decay $B^+\to~D(K_{\rm S}^0\pi^+\pi^-\pi^0)K^+$ to determine $\phi_3$ using the same formalism as with $B^{+}\to D(K_{\rm S}^0\pi^+\pi^-)K^{+}$~\cite{GGSZ, GGSZ2}. The decay $D\to K_{\rm S}^{0}\pi^{+}\pi^{-}\pi^{0}$ is a suitable addition because it has a branching fraction of 5.2$\%$~\cite{PDG}, which is large compared to that of other multibody final states including $K_{\rm S}^0\pi^{+}\pi^{-}$. The decay occurs through many intermediate resonances, such as $K_{\rm S}^{0}\omega$ and $K^{*\pm}\rho^{\mp}$, that lead to variations of the strong-phase difference over the phase space, a requirement for extracting $\phi_3$ from a single final state. However, a significant complication is that the four-body final state requires a binning of the five-dimensional $D$ phase space, as opposed to a two-dimensional Dalitz plot for the three-body case.  The measurement is performed with an $e^{+}e^{-}$ collision data sample collected by the Belle detector at a centre-of-mass energy corresponding to the $\Upsilon(4S)$ resonance. The sample contains $772\times 10^{6}$ $B\overline{B}$ events corresponding to an integrated luminosity of 711~fb$^{-1}$.  As an input to the analysis, we use the strong-phase difference measurements in phase space bins~\cite{Resmi} obtained from an analysis of CLEO-c~\cite{CLEO1,CLEO2,CLEO3,CLEO4} data~\footnote{Normal activities of the CLEO Collaboration ceased in 2012. However, access to the data was still possible for former CLEO Collaboration members and several results have been published. Any such publication, such as ref.~\cite{Resmi} are not official results of the CLEO Collaboration. Hence we refer to results from CLEO-c data rather than from the CLEO Collaboration.}. 

The remainder of this paper is arranged as follows. Section~\ref{sec:GGSZ} describes the formalism of the method for measuring $\phi_3$. The inputs derived from CLEO-c data and the Belle data and detector are described in sections~\ref{Sec:CLEO}  and \ref{Sec:datadetector}, respectively, after which an overview of the analysis strategy is presented, in section~\ref{Sec:ana}. The event selection criteria are given in section~\ref{Sec:sel}, and the signal yield determination in the flavour-tagged $D$ sample, which is a required input to the analysis, is presented in section~\ref{Sec:Dst}.  The measurement of $CP$ violation in the $B$ sample in bins of the $D$ phase space is explained in section~\ref{Sec:BtoDh} and the related systematic uncertainty estimation is listed in section~\ref{Sec:syst}. The extraction of the physics parameter $\phi_3$ and the average of this result with previous Belle measurements are presented in section~\ref{Sec:phi3}, before conclusions given in section~\ref{Sec:con}.

\section{Formalism for measuring $\phi_3$ with  $B^{+}\to D(K_{\rm S}^0\pi^+\pi^-\pi^0)K^{+}$ decays }
\label{sec:GGSZ}

The determination of $\phi_3$ from $B^{+}\to D K^{+}$ decays, where the $D$ meson decays to a multibody self-conjugate final state, can be performed via two methods: model-dependent and -independent. The model-dependent method requires a model of the amplitudes describing the intermediate resonances and partial waves, assumed to be contributing to the decay, to be fitted to the distribution of events over the $D$ phase space. Model assumptions used in the method lead to a difficult determination of systematic uncertainty and may limit the precision of the $\phi_3$ measurement, to as much as $\pm$8\textendash 9$^{\circ}$~\cite{Belle-ModelDep}. On the other hand, the model-independent method requires that measurements of $CP$-violating asymmetries are made in distinct regions of $D$ meson phase space, which we refer to as bins. The binning reduces the statistical precision compared to the model-dependent method, but the uncertainty related to model assumptions is removed by using independent measurements of the average strong-phase differences, bin-by-bin.
The average strong-phase differences can be determined using $e^+e^-$ collision data at the open-charm threshold, which has been done for  $D^0\to K_{\rm S}^0\pi^+\pi^-\pi^0$~\cite{Resmi}. Therefore, given its systematic robustness, we follow the model-independent approach. We introduce the method in the remainder of this section.

The amplitude  for the decay $B^{+}\to DK^{+}$, $D\to f$, where $f$ is a common multibody final state from the $D^0$ and $\overline{D}^0$ decay, can be written as
\begin{equation}
A_{B} = \overline{A} + r_{B}e^{i(\delta_{B}+\phi_{3})}A, 
\end{equation}
where $A$ is the amplitude for $D^{0}\to f$ at a point in the allowed phase space $\mathcal{D}$, $\overline{A}$ is the amplitude for $\overline{D}^0\to f$ at the same point in phase space, $r_{B}$ is the ratio of the absolute values of $B^+\to \overline{D}^0K^+$ and $B^+\to D^0K^+$ decay amplitudes, and $\delta_B$ is the strong-phase difference between the two $B\to DK$ amplitudes. Thus, the probability density for a decay at a point in $\mathcal{D}$ is 
\begin{align}
P_{B} = |A_{B}|^{2} & = |\overline{A}|^{2} + r_{B}^{2}|A|^{2} +2 r_{B}\Re\left[\overline{A}^{*}Ae^{i(\delta_{B}+\phi_{3})}\right]. \label{eq:pB}
\end{align}
Furthermore,
\begin{equation}
\overline{A}^{*}A = |\overline{A}||A|e^{i(\delta_{D}-\delta_{\overline{D}})} = |\overline{A}||A|e^{i\Delta \delta_{D}},
\end{equation}
where $\delta_{D}$ and $\delta_{\overline{D}}$ are the strong phases for $D^{0}\to f$ and $\overline{D}^0\to f$ decays, respectively. With this,  eq.~\ref{eq:pB} becomes
\begin{eqnarray}
P_{B} &=  &|\overline{A}|^{2} + r_{B}^{2}|A|^{2} + 2r_{B}|\overline{A}||A|\left[\cos\Delta \delta_{D} \cos(\delta_{B}+\phi_{3}) - \sin\Delta \delta_{D} \sin(\delta_{B}+\phi_{3})\right] \nonumber \\
 &= & \overline{P} + r_{B}^{2}P + 2\sqrt{P\overline{P}}(x_{+}C - y_{+}S),
\label{Eq:pB1}
\end{eqnarray}
where $P=|A|^2$, $\overline{P} = |\overline{A}|^{2}$, $x_{+} = r_{B}\cos(\delta_{B}+\phi_{3})$, $y_{+} = r_{B}\sin(\delta_{B}+\phi_{3})$, $C = \cos\Delta \delta_{D}$ and $S = \sin\Delta \delta_{D}$. For the charge-conjugate mode, $B^{-}\to DK^{-}$, the density is given by the same expression, with $A\leftrightarrow \overline{A}$ and $\phi_3 \to -\phi_3$, which leads to the definitions $x_{-} = r_{B}\cos(\delta_{B}-\phi_{3})$ and  $y_{-} = r_{B}\sin(\delta_{B}-\phi_{3})$. The partial decay rates for $B^{\pm}\to DK^{\pm}$ within the $i^{\rm th}$ bin of $\mathcal{D}$, which corresponds to a subset of phase space $\mathcal{D}_i$, are
\begin{equation}
\Gamma_{i}^{-} = h(K_{i} + r_{B}^{2}\overline{K}_i +2\sqrt{K_{i}\overline{K}_i}(c_{i}x_{-} + s_{i}y_{-})),  \label{Eq:B-}
\end{equation}
\begin{equation}
\Gamma_{i}^{+} = h(\overline{K}_i + r_{B}^{2}K_{i} +2\sqrt{K_{i}\overline{K}_i}(c_{i}x_{+} - s_{i}y_{+})), \label{Eq:B+}
\end{equation} 
where $K_{i}$ and $\overline{K}_i$ are the fractions of flavour-tagged $D^{0}$ and $\overline{D}^0$ events in $\mathcal{D}_i$ and $h$ is the common normalization factor. A sample of $D^{0}\to K_{\rm S}^0\pi^+\pi^-\pi^0$ candidates from $D^{*+}\to D^0\pi^+$ decays, where the charge of the pion from the $D^{*+}$ decay tags the flavour of the $D$ meson, are used to determine values of $K_{i}$ and $\overline{K}_i$. The $c_i$ and $s_i$ parameters are the amplitude-weighted averages of $C$ and $S$ over the region $\mathcal{D}_i$. The $c_i$ parameter is defined as
\begin{equation}
c_i = \frac{\int_{\mathcal{D}_i}\sqrt{P\overline{P}}C\:d\mathcal{D}}{\sqrt{\int_{\mathcal{D}_i} P\:d\mathcal{D}\int_{\mathcal{D}_i}\overline{P}\:d\mathcal{D}}},
\end{equation}
 and the definition of $s_i$ is the same, with $C$ being replaced by $S$. Therefore, with $c_{i},~s_{i},~K_{i}$, and $\overline{K}_i$ given as external inputs to the analysis, one can determine $x_{\pm}$, $y_{\pm}$ and $h$ from a set of partial decay rate measurements, when $\mathcal{D}$ is divided into three or more bins. The loss of statistical precision can be mitigated by increasing the number of bins; with an increased number of bins, however, the uncertainty on the external inputs also increases, limiting the precision of the measurement. The method by which the values of $x_{\pm}$ and $y_{\pm}$ are used to constrain $\phi_3$, $r_B$ and $\delta_B$ is described in section~\ref{Sec:phi3}.

\section{External measurements of $c_i$ and $s_i$}
\label{Sec:CLEO}

The values of $c_i$ and $s_i$ for the decay $D\to K_{\rm S}^{0}\pi^{+}\pi^{-}\pi^{0}$ have been determined using $e^{+}e^{-}$ collision data collected at a centre-of-mass energy corresponding to the $\psi(3770)$ resonance~\cite{Resmi}. The quantum correlations between the neutral $D$ mesons produced in decays of the $\psi(3770)$ are exploited to extract the strong-phase differences in bins of the phase space. This four-body final state has a five-dimensional phase space, which was divided into nine exclusive bins, selected to contain different intermediate resonances, thus minimizing the strong-phase variation within the bin as much as possible. The sensitivity of the binning could be improved upon using an amplitude model of $D^0\to K_{\rm S}^0\pi^{+}\pi^{-}\pi^{0}$, which is unavailable at present. The binning scheme is listed in table~\ref{Table:Bin}. In each successive bin, only events that do not belong to the previous bins are selected (e.g. bin 2 is populated by events with $m_{K_{\rm S}^0\pi^-}$ and $m_{\pi^+\pi^0}$ within the denoted intervals, and $m_{\pi^+\pi^-\pi^0}$ not in the denoted interval for bin 1). The bins are thus exclusive.

\begin{table}[t] 
\centering
\small{
\begin{tabular} {l c c c}
\hline 
 Bin no. & Bin region & $m_{\rm L}$   &$m_{\rm U}$     \\[0.5ex]
 & & (GeV/$c^{2}$) & (GeV/$c^{2}$) \\[0.5ex]
\hline
\hline
 1 & m$_{\pi^{+}\pi^{-}\pi^{0}}$ $\approx$ m$_{\omega}$ & 0.762 & 0.802\\[0.5ex]
 2 & m$_{K_{\rm S}^{0}\pi^{-}}$ $\approx$ m$_{K^{*-}}$  & 0.790 & 0.994 \\[0.5ex]
  &  m$_{\pi^{+}\pi^{0}}$ $\approx$ m$_{\rho^{+}}$ & 0.610 & 0.960 \\[0.5ex]
 3 & m$_{K_{\rm S}^{0}\pi^{+}}$ $\approx$ m$_{K^{*+}}$  & 0.790 & 0.994 \\[0.5ex]
  & m$_{\pi^{-}\pi^{0}}$ $\approx$ m$_{\rho^{-}}$ & 0.610 & 0.960\\[0.5ex]
 4 & m$_{K_{\rm S}^{0}\pi^{-}}$ $\approx$ m$_{K^{*-}}$ & 0.790 & 0.994\\[0.5ex]
 5 & m$_{K_{\rm S}^{0}\pi^{+}}$ $\approx$ m$_{K^{*+}}$ & 0.790 & 0.994\\[0.5ex]
 6 & m$_{K_{\rm S}^{0}\pi^{0}}$ $\approx$ m$_{K^{*0}}$ & 0.790 & 0.994\\[0.5ex]
 7 & m$_{\pi^{+}\pi^{0}}$  $\approx$ m$_{\rho^{+}}$ & 0.610 & 0.960 \\[0.5ex]
 8 & m$_{\pi^{-}\pi^{0}}$  $\approx$ m$_{\rho^{-}}$& 0.610 & 0.960\\[0.5ex]
 9 & Remainder & - & -\\[0.5ex]
\hline
\end{tabular} 
\caption{Specifications of the nine exclusive bins of $D\to K_{\rm S}^{0}\pi^{+}\pi^{-}\pi^{0}$ phase space. Here  $m_{\rm L}$ and $m_{\rm U}$ are the lower and upper limit, respectively, of the invariant mass in each region~\cite{Resmi}. }\label{Table:Bin}
}
\end{table}  

\begin{figure}[t]
\centering
\includegraphics[width=0.75\columnwidth]{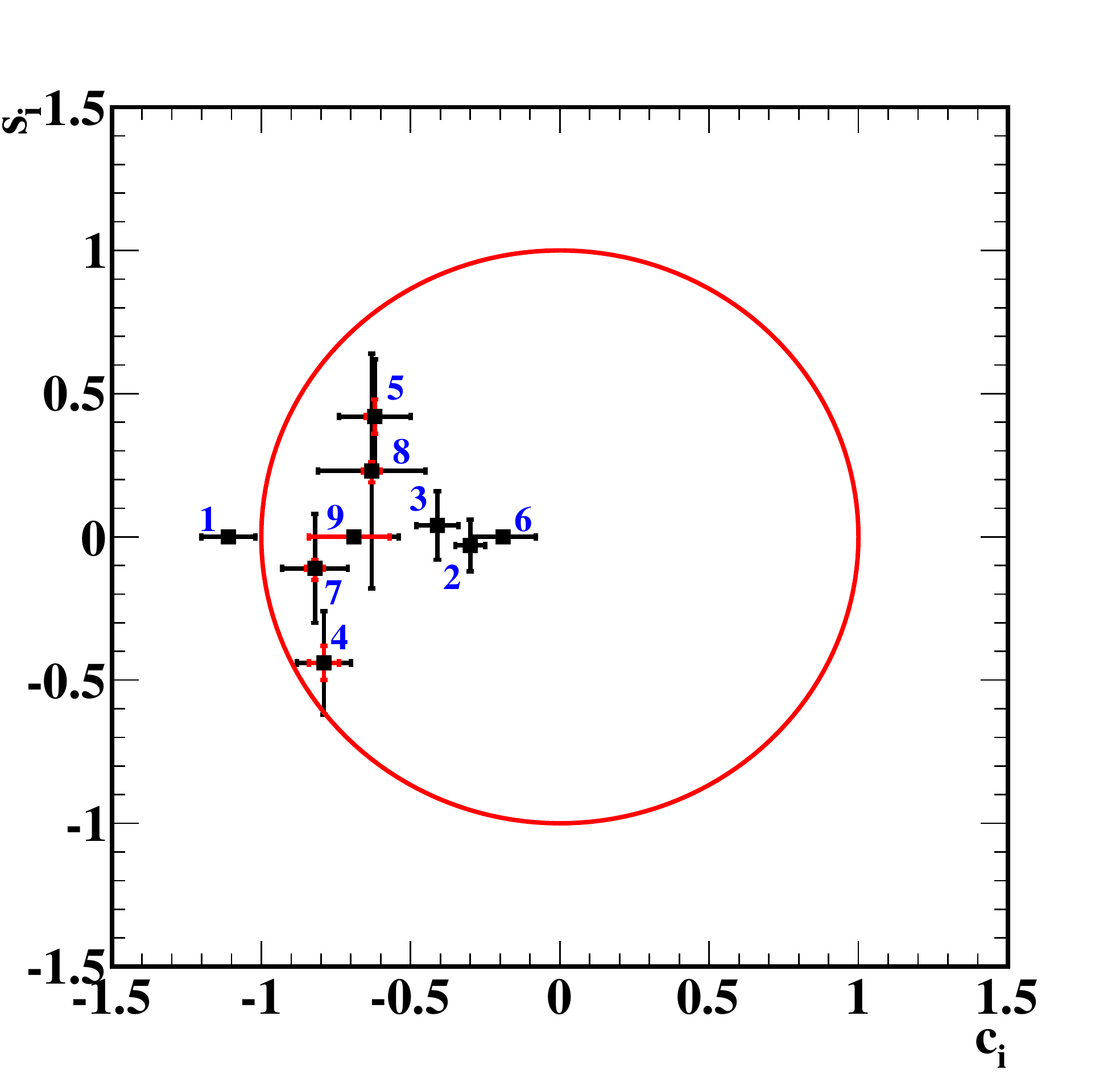}
\caption{Values of $c_{i}$ and $s_{i}$ reported in ref.~\cite{Resmi}. The black and red error bars represent statistical and systematic uncertainties, respectively.}\label{Fig:cisi}
\end{figure}

Certain constraints are imposed in the fit, which arise from the nature of the symmetry between the bins, to extract $c_i$ and $s_i$ parameters. Bins 1, 6 and 9 are $CP$ self-conjugate, which implies
\begin{equation}
s_{1} = 0,~ s_{6} = 0,~ s_{9}=0. \label{Eqn:s1s6s9}
\end{equation}
Bin 9 is $CP$ self-conjugate because the region corresponding to the sum of bins 1 to 8 is $CP$ self-conjugate. Bins 2 and 3, 4 and 5, and 7 and 8 are pairwise $CP$-conjugate, which imposes a relation between their $s_{i}$ values:
\begin{equation}
s_{i}\sqrt{K_{i}\overline{K}_i} + s_{i+1}\sqrt{K_{i+1}\overline{K}_{i+1}} = 0, \label{Eqn:s2s3}
\end{equation}
where $i =$ 2, 4 and 7. The results for $c_{i}$ and $s_{i}$ are summarized in table~\ref{Table:cisi} and are shown in figure~\ref{Fig:cisi}. In the analysis we use the same binning scheme so that $c_i$ and $s_i$ can be taken as external inputs in determining $x_{\pm}$ and $y_{\pm}$.

\begin{table} [t] 
\centering
\begin{tabular} {l  c c} 
\hline 
Bin no. &  $c_{i}$ & $s_{i}$\\[0.5ex]
\hline
\hline
1 &         $-1.11\pm0.09_{-0.01}^{+0.02}$ & 0.00\\[0.5ex]
2 &     $-0.30\pm 0.05 \pm 0.01$ & $-0.03\pm 0.09_{-0.02}^{+0.01}$\\[0.5ex]
3 &         $ -0.41\pm0.07_{-0.01}^{+0.02}$ & $0.04\pm0.12_{-0.02}^{+0.01~*}$\\[0.5ex]
4 &         $-0.79\pm0.09\pm 0.05$ &$-0.44\pm0.18\pm0.06$\\[0.5ex]
5 &           $-0.62\pm0.12_{-0.02}^{+0.03}$ & $0.42\pm0.20\pm0.06^{~*}$ \\[0.5ex]
6 &          $-0.19\pm0.11\pm 0.02$ & 0.00\\[0.5ex]
7 &           $-0.82\pm0.11\pm 0.03$ & $-0.11\pm0.19_{-0.03}^{+0.04}$\\[0.5ex]
8 &           $-0.63\pm0.18\pm 0.03$& $0.23\pm0.41_{-0.03}^{+0.04~*}$ \\[0.5ex]
9 &           $-0.69\pm0.15_{-0.12}^{+0.15}$ & 0.00\\[0.5ex]
\hline
\end{tabular}
\caption{Values of $c_{i}$ and $s_{i}$ reported in ref.~\cite{Resmi}. The uncertainties are statistical and systematic, respectively. The $s_{i}$ results marked by * in bins 3, 5 and 8 are derived from those in other bins, according to the constraints of eq.~\eqref{Eqn:s2s3}. The statistical uncertainty on these $s_i$ values include contribution from $K_i$ values according to the error propagation formalism.}\label{Table:cisi}
\end{table}

\section{Data samples and the Belle detector}
\label{Sec:datadetector}
We use an $e^{+}e^{-}$ collision data sample containing $772\times10^6~B\overline{B}$ events collected by the Belle detector at a centre-of-mass energy corresponding to the pole of the $\Upsilon(4S)$ resonance. Monte Carlo (MC) simulated samples are used to optimize the selection, determine selection efficiencies, and identify sources of background. The MC samples of signal and background processes are generated using {\tt EvtGen}~\cite{Evtgen} with the {\tt GEANT}~\cite{Geant} package being subsequently used to model the detector response to the decay products. {\tt PHOTOS}~\cite{PHOTOS} incorporates effects due to final-state radiation from charged particles.

The Belle detector~\cite{Belle1,Belle2} was located at the interaction point of the KEKB asymmetric-energy $e^+e^-$ collider~\cite{KEKB,KEKB2}. The detector subsystems relevant for this study are: the silicon vertex detector (SVD) and central drift chamber (CDC), for charged particle tracking and measurement of energy loss due to ionization ($dE/dx$); the aerogel threshold Cherenkov counters (ACC) and time-of-flight (TOF) scintillation counters, for particle identification~(PID); and the electromagnetic calorimeter (ECL) consisting of an array of CsI(Tl) crystals to measure photon energies. These subsystems are situated in a magnetic field of 1.5~T. A more detailed description of the Belle detector can be found in refs.~\cite{Belle1,Belle2}.

\section{Analysis Overview}
\label{Sec:ana}

The essence of the analysis lies in eqs.~\eqref{Eq:B-} and \eqref{Eq:B+}, which describe the partial decay rates in each bin. However, these relations do not account for experimental resolution and acceptance. For example, the invariant mass resolution causes events to be assigned to bins outside of their origin, an effect we shall call \enquote{migration}.  The background contributions are to be considered as well. Here we briefly summarize how these experimental effects are accounted for.

 \subsection{Efficiency}
 Three different samples are used in this analysis, each with differing selection efficiencies due to the kinematic differences between the final states: the quantum correlated $D\overline{D}$ sample from $\psi(3770)$ decays, the Belle sample of $B^{+}\to Dh^{+}$, where $h = (K, \pi)$, and the Belle sample of $D^{*+}\to D\pi^{+}$ used to determine $K_i$ and $\overline{K}_i$. The sample of $B^{+}\to D\pi^{+}$ is used as a control sample, as it is topologically identical to the signal, but with negligible expected $CP$ violation~\cite{Dpi}. The $c_i$ and $s_i$ results measured with CLEO-c data have been corrected for efficiency. Efficiency variation among bins will not matter if the efficiency profile is the same for both $B^{+}\to Dh^{+}$ and flavour-tagged $D$ samples. This is partially achieved by requiring similar kinematic properties for the $D$ meson in both samples. The efficiency profile depends primarily on $D$ momentum, hence we select the flavour-tagged $D$ sample in such a way that the $D$ momentum approximately matches that of the $B^{+}\to Dh^{+}$ sample. The matching is not exact, so independent efficiency corrections are applied to the yields in both samples while calculating the parameters of interest. 
 \subsection{Momentum resolution}
 The finite momentum resolution causes events to migrate among the bins. The $c_i$ and $s_i$  results are obtained after applying corrections for these migration effects. The amount of migration in both $B$ and $D^*$ samples is estimated as a migration matrix $M_{ij}$. The matrix has its diagonal elements close to one, and off-diagonal elements are small. MC samples of signal events are used to obtain the migration matrix. The data yield in each bin, $Y_i$, is modified as $Y_{i}' = M_{ij}Y_j$. 
 Any difference  between the invariant mass resolution in the data and MC samples must be taken into account. We find the effect of the difference in resolution is only significant in bin 1, which contains the $\omega$ resonance. This bin is narrow due to the small natural width of the $\omega$. However, the natural width is the same order as the $m_{\pi^+\pi^-\pi^0}$ resolution, so there is significant migration out of this bin that is not compensated by migration into bin 1. Therefore, the  $M_{1j}$ elements of the migration matrix are determined after applying a Gaussian smearing to the value of $m_{\pi^+\pi^-\pi^0}$ by a scale factor. The scale factor is obtained from the observed difference in $\omega$ mass resolution between data and MC samples. The scale factors are 1.13~$\pm$~0.02 and 1.09~$\pm$~0.02 for the $B^+\to Dh^+$ and $D^{*+}\to D\pi^+$ samples, respectively.
 \subsection{Signal extraction}
It is important to account for the background contributions in the sample while extracting the specified parameters. An extended maximum likelihood fit is performed on the data in each bin of the flavour-tagged $D$ sample to extract the values of $K_i$ and $\overline{K}_i$. The fit to the $B$ sample in the bins of $D$ phase space is performed using an extended likelihood fit that simultaneously fits all bins in the $B^{+}\to DK^+$ and $B^{+}\to D\pi^+$ decay modes, so that the values of the parameters $x_{\pm}$ and $y_{\pm}$ that are common to the expectation for each bin yield, can be extracted, as well as the cross-feed between these samples.

\section{Event selection}
\label{Sec:sel}

We reconstruct the decays $B^{+}\to DK^{+}$ and $B^{+}\to D\pi^{+}$, where the neutral $D$ meson decays to the four-body final state of $K_{\rm S}^0\pi^+\pi^-\pi^0$. In addition, $D^{*+} \to D^0\pi^{+}$ decays produced via the $e^+e^- \to c\overline{c}$ continuum process are selected to determine the $K_i$ and $\overline{K}_i$ parameters.

For charged particle candidates originating directly from the $B$ and $D$ decays, we require that the track be within 0.5~cm and  $\pm$3.0~cm of the interaction point~(IP) in the directions perpendicular to~(radial) and parallel to the $z$-axis, respectively; the $z$-axis is defined to be opposite to the $e^+$ beam direction. The charged tracks are classified as pions or kaons based on information from CDC, ACC, and TOF sub-detector systems. The pion (kaon) identification efficiency is 92\% (84\%) and the probability of misidentification as a kaon (pion) is 15\% (8\%) \cite{bib:horiisan}.

We select $K_{\rm S}^0$ candidates from two oppositely charged tracks assumed to be pions.
The invariant mass of the two tracks is required to be within the range 0.487\textendash 0.508~GeV/$c^2$ corresponding to $\pm 3\sigma$ of the known $K_{\rm S}^0$ mass~\cite{PDG}, where $\sigma$ is the mass resolution. A neural network~\cite{NB} based selection is applied on the daughter tracks to remove background from random combinations~\cite{nisks}. The input variables to the neural network are the $K_{\rm S}^0$ momentum in the lab frame, the distance between the two track helices along the $z$-axis at their point of closest approach, the $K_{\rm S}^0$ flight length in the radial direction, the angle between the $K_{\rm S}^0$ momentum and the vector joining the IP to the $K_{\rm S}^0$ decay vertex, the angle between pion momentum and the boost direction of lab frame in $K_{\rm S}^0$ rest frame and pion momentum in $K_{\rm S}^0$ rest frame, the distances of closest approach in the radial direction between IP and the two pion helices,  the number of hits in CDC for each pion track, and the presence of hits in the SVD for each pion track. The $K_{\rm S}^0$ selection efficiency is 87\%, which is determined from an MC sample of generic $B\overline{B}$ events.  

The $\pi^{0}$ candidates are reconstructed from pairs of photons detected in the ECL. We select candidates with diphoton invariant mass $ M_{\pi^0}$ in the range 0.119\textendash 0.148~GeV/$c^{2}$, which corresponds to $3\sigma$ about the nominal $\pi^{0}$ mass~\cite{PDG}. The photon energy thresholds are optimized separately for $\pi^0$ candidates detected in combinations of the barrel, forward endcap (FWD EC) and backward endcap (BWD EC) regions of the ECL as given in table~\ref{Table:OptEg} by maximizing the significance $S/\sqrt{S+B}$, where $S$ and $B$ are the number of signal and background events selected from MC samples in the signal region, respectively. (The criteria that define the signal region are described later in this section.) 
\begin{table} [t] 
  \centering
    \begin{tabular} {ll c c }
   \hline 
 $\gamma_{1}$ & $\gamma_{2}$ & $E_{\gamma_1}$ (MeV)& $E_{\gamma_2}$ (MeV)\\[0.5ex]
 \hline
 \hline
  Barrel & Barrel & $\phantom{0}$70 & $\phantom{0}$65\\[0.5ex]
  FWD EC & Barrel & 220 & $\phantom{0}$65\\[0.5ex]
  Barrel & BWD EC & $\phantom{0}$65 & $\phantom{0}$95 \\[0.5ex]
  FWD EC & FWD EC & 150 & 210 \\[0.5ex]
    \hline
 \end{tabular} 
 \caption{Optimized $E_{\gamma}$ thresholds for the photon candidates. The FWD EC, barrel, and BWD EC regions of the ECL are defined in the polar angle ranges (12.4$^{\circ}$, 31.4$^{\circ}$), (32.2$^{\circ}$, 128.7$^{\circ}$), and (130.7$^{\circ}$, 155.1$^{\circ}$), respectively. 
 }\label{Table:OptEg}
   \end{table} 
Studies of MC samples indicate that candidates in the other ECL sector combinations make up only  1.5\% of the total, and a common energy threshold of 50 MeV is applied on these. All selected combinations of $K_{\rm S}^{0}\pi^{+}\pi^{-}\pi^{0}$ candidates are retained for further study. In addition, kinematic constraints are applied to the $K_{\rm S}^0$, $\pi^0$, and $D$ invariant masses and decay vertices to improve the momentum resolution of the $B$ candidates, as well as the invariant masses used to bin the $D$ phase space.

\begin{figure}[t]
\centering
\includegraphics[width=0.5\columnwidth]{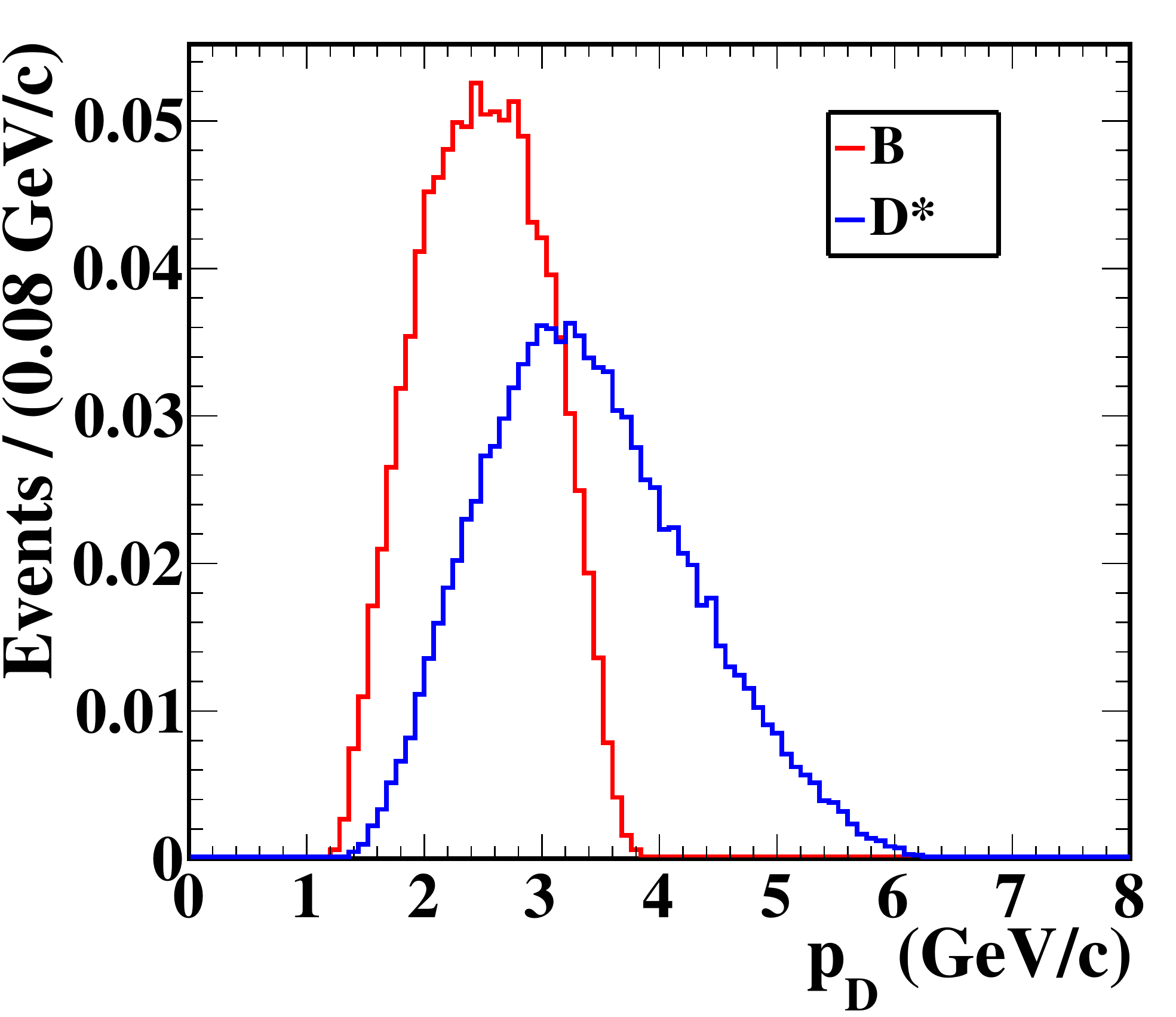}
\caption{Distributions of the measured $D$ meson momenta $p_D$ in the lab frame for (blue) $D^{*+}$ and (red) $B^+ \to Dh^+$  signal MC samples.  }\label{Fig:pD}
\end{figure}

The $D^{*+}\to D^0\pi^{+}$ decay uses the charge of the accompanying pion to identify the flavour of the $D$ meson. This pion is referred to as a slow pion because of the limited phase space of the decay that results in it having lower momentum on average than other final-state particles. To improve the momentum resolution of the slow pion, it is required to have at least one hit in the SVD. Signal $D^{*+}$ candidates are identified by two kinematic variables: $M_{D}$, the invariant mass of the $D$ candidate, and $\Delta M$, the difference in the invariant masses of $D^{*+}$ and $D$ meson candidates. The events that satisfy the criteria, $1.80 <M_D < 1.95 $~GeV/$c^2$ and $\Delta M < 0.15$~GeV/$c^2$ are retained. The $D$ meson momentum in the lab frame is chosen to be in the range 1\textendash 4~GeV/$c$ to approximately match the range of $D$ momentum in the $B^{+}\to Dh^{+}$ sample, as illustrated in figure~\ref{Fig:pD}.

The $D$ and $\pi^{+}$ candidates are constrained to come from a common vertex to form the $D^{*+}$ candidate. On average, there are 1.6 $D^{*+}$ candidates in an event. If there is more than one candidate in an event, the candidate with the smallest $\chi^2$ value from the $D^{*+}$ vertex fit is retained. This criterion selects the correct signal candidate in 69\% of the events with multiple candidates. The overall selection efficiency is 3.7\%, which includes the secondary branching fraction of $K_{\rm S}^0 \to \pi^+\pi^-$.

A $D$ candidate is combined with a charged kaon (pion) track to form a $B^{+}\to DK^{+}$ ($B^{+} \to D\pi^{+}$) candidate. The invariant mass of the $D$ candidate is required to be in the range 1.835\textendash 1.890~GeV/$c^2$. The signal candidates are identified using two kinematic variables, the energy difference $\Delta E$ and beam-energy-constrained mass $M_{\rm bc}$, which are  defined as
$\Delta E = E_{B} - E_{\rm beam}$ and $M_{\rm bc} = c^{-2}\sqrt{E_{\rm beam}^{2}-|\vec{\mathbf{p}}_{B}|^2c^{2}}$, where $E_{B}$ $(\vec{\mathbf{p}}_{B})$ is the energy (momentum) of the $B$ candidate and $E_{\rm beam}$ is the beam energy in the centre-of-mass frame . We select candidates that satisfy the criteria $M_{\rm bc}> 5.27$~GeV/$c^2$ and $-0.13 < \Delta E < 0.30$~GeV. The asymmetric $\Delta E$ window is chosen to avoid the peaking structure appearing at lower values from partially reconstructed $B^+ \to D^{(*)}K^{(*)+}$ decays. The signal region used while performing optimization of the selection is $|\Delta E| < 0.05$~GeV. The average $B$ candidate multiplicity is  1.3. In events with more than one candidate, we retain the candidate with the smallest value of $(\frac{M_{\rm bc}-M_{B}^{\rm PDG}}{\sigma_{M_{\rm bc}}})^{2} + (\frac{M_{D}-M_{D}^{\rm PDG}}{\sigma_{M_{D}}})^{2} + (\frac{M_{\pi^{0}}-M_{\pi^{0}}^{\rm PDG}}{\sigma_{M_{\pi^{0}}}})^{2} $. Here, the masses $M_i^{\rm PDG}$ are those reported by the Particle Data Group~\cite{PDG} and the resolutions $\sigma_{M_{\rm bc}},~\sigma_{M_{D}}$, and $\sigma_{M_{\pi^{0}}}$ are obtained from MC simulated samples of signal events. The best candidate selection criterion is 80\% efficient in selecting the correctly reconstructed candidate.

The background from $e^+e^- \to q\overline{q}~ (q = u, d, s, c)$ continuum processes is suppressed by exploiting the difference in event topology compared to $e^+e^- \to \Upsilon (4S) \to B\overline{B}$ events. The continuum events are jet-like in nature, whereas $B\overline{B}$ events have a spherical topology, due to the low momentum of the $B$ mesons produced via the $\Upsilon(4S)$ resonance. 
A neural-network-based algorithm~\cite{NB} is used to discriminate between continuum background and $B$ events. We also use variables related to the displaced vertices of $B$ decays from the IP and the associated leptons/kaons from the non-signal $B$ meson in the event, which give an additional handle to distinguish continuum events. 

The eight input variables to the neural network are the likelihood ratio obtained via Fisher discriminants~\cite{Fisher} formed from modified Fox-Wolfram moments~\cite{KSFW1,KSFW2}, the absolute value of the cosine of the angle between the $B$ candidate and the $z$ axis in the $e^{+}e^{-}$ centre-of-mass frame, the absolute value of the cosine of the angle between the thrust axis of the $B$ candidate and that of the rest of the event in the centre-of-mass frame, the vertex separation between the two $B$ candidates~\cite{BVertex} along $z$-axis, the absolute value of the $B$ flavour-dilution factor~\cite{FlavorTag}, the difference between the sum of the charges of particles in the hemisphere about the $D$ direction in the centre-of-mass frame and the one in the opposite hemisphere, excluding the particles used for the reconstruction of $B$, the product of the charge of the $B$ and the sum of the charges of all kaons not used for reconstruction of $B$, and the cosine of the angle between the $D$ direction and the direction opposite to that of the $\Upsilon(4S)$ in the $B$ rest frame. 

Signal and continuum MC samples  are used to train the neural network. We require the neural network output, $C_{\rm NN}$, to be greater than $-0.6$, which reduces the continuum background by 67\% with a loss of only 5\% of the signal. The overall selection efficiency is 4.7\% and 5.3\% for $B^{+}\to DK^{+}$ and $B^{+}\to D\pi^{+}$ decays, respectively. These efficiencies include the secondary branching fraction of $K_{\rm S}^0 \to \pi^+\pi^-$. The efficiencies in each bin and the migration matrix for the $B^+\to Dh^+$ selection are given in appendix~\ref{app:migmatrix}.

\section{Determination of $K_i$ and $\overline{K}_i$ from the $D^{*+}$ sample}
\label{Sec:Dst}

The fractions of $D^0$ and $\overline{D}^0$ events in each $D$ phase space bin, represented as $K_i$ and $\overline{K}_i$, are measured from the selected sample of $D^{*+} \to D\pi^{+}$ candidates. The yield of signal events is obtained from a two-dimensional unbinned extended maximum-likelihood fit to the distribution of $M_D$ and $\Delta M$ for the selected candidates. The fit is performed independently in each bin. In general, there are two types of background: {\it combinatorial}, which is due to the random combination of final-state particles to form a $D^{*+}$ candidate, and {\it random-slow-pion}, in which a correctly reconstructed $D$ meson combines with a $\pi^{+}$, which is not from a common $D^{*+}$ decay, to form a candidate. The combinatorial background peaks neither in the $M_{D}$ nor $\Delta M$ distributions, whereas the random-slow-pion background peaks only in the $M_{D}$ distribution.

The signal component of the $M_{D}$ distribution is described by a probability density function (PDF) that is the sum of a Crystal Ball (CB)~\cite{CB} function and two Gaussian functions with a common mean. The combinatorial background PDF is parametrized by a first-order polynomial. The signal PDF is also used to model the random-slow-pion background distribution in $M_{D}$. The $\Delta M$ signal PDF is described by the sum of an asymmetric Gaussian and three Gaussian functions with a common mean. The combinatorial background  $\Delta M$  distribution is parametrized by the sum of a threshold function and two Gaussian PDFs. The threshold function is
 \begin{equation}
 				f(\Delta M) =  (\Delta M-m_{\pi})^{\frac{1}{2}} + \alpha (\Delta M-m_{\pi})^{\frac{3}{2}} + \beta (\Delta M-m_{\pi})^{\frac{5}{2}}, \label{Eq:Threshold}
 \end{equation} 
where $m_{\pi}$ is the mass of a charged pion~\cite{PDG}, and $\alpha$ and $\beta$ are shape parameters. In the final fit to data, the shape parameters are fixed to the values obtained from MC. The Gaussian functions describe a small peak in the $\Delta M$ combinatorial distribution, which is due to misreconstructed $\pi^0$ candidates. The parameters of the Gaussian functions and the fraction of candidates in the peak are fixed to the values obtained from a MC sample. The random-slow-pion background PDF is the same as the threshold function used to describe the combinatorial background.

The signal $M_D$ and $\Delta M$  PDFs  are correlated such that the width of the $\Delta M$ distribution depends upon $M_D$. The width of the core Gaussian in the $\Delta M$ signal PDF is parametrized as 
\begin{equation}
\sigma(\Delta M) = a_{0} + a_{2} (M_{D} - M_{D}^{\rm PDG})^{2}, \label{Eq:Corre}
\end{equation}
where $a_0$ and $a_2$ are parameters to be determined from data. The correlation between $M_D$ and $\Delta M$ distributions is found to be negligible in studies of background MC samples. Therefore, the one-dimensional PDFs are multiplied to obtain the total background PDF.

\begin{figure}[t]
\centering
\begin{tabular}{cc}
\includegraphics[width=0.5\columnwidth]{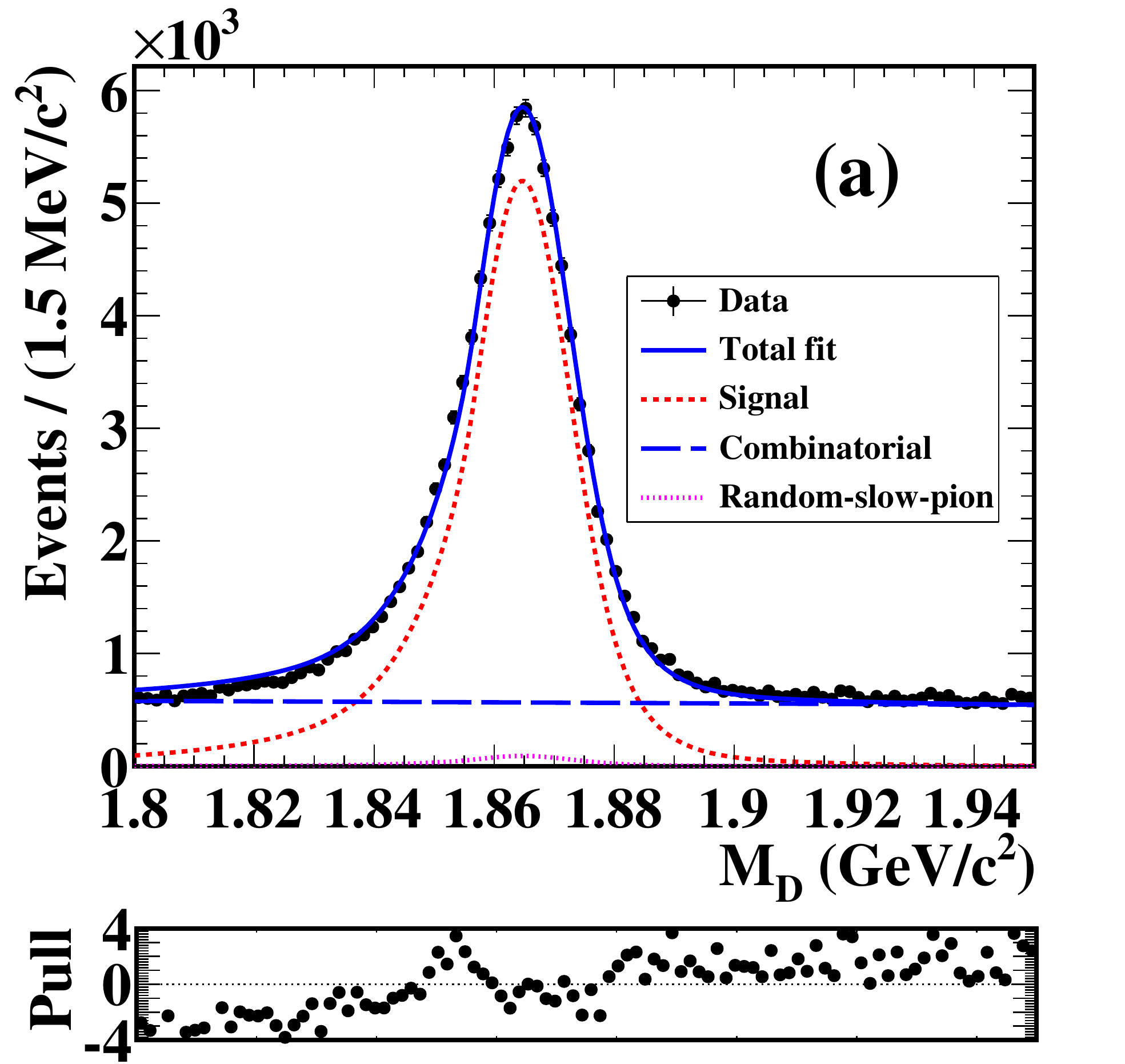}&
\includegraphics[width=0.5\columnwidth]{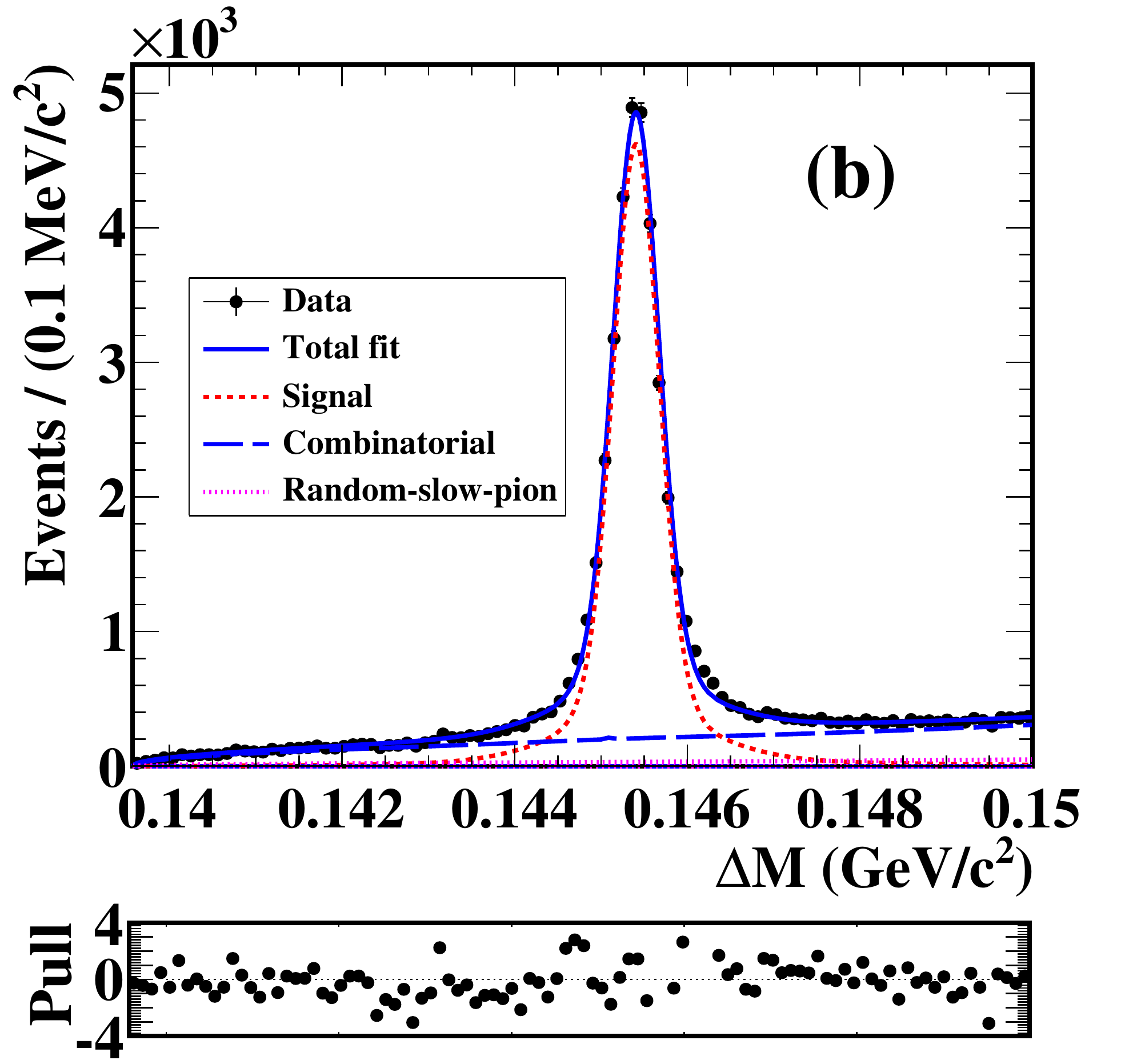}\\
\end{tabular}
\caption{Signal-enhanced fit projections of (a) $M_{D}$  and (b) $\Delta M$ distributions from $D^{*\pm} \to D\pi^{\pm}$ data sample in bin 1. The black points with error bars are the data and the solid blue curves show the total fit. The error bars are barely visible as they are smaller than the size of the points. The dotted red, blue and magenta curves represent the signal, combinatorial and random-slow-pion backgrounds, respectively. The pull between the fit and the data is shown below the distributions.}
\label{Fig:Dst_total}
\end{figure}

The yields, except that describing the peaking component in the combinatorial background $\Delta M$ distribution and the shape parameters $a_{0(2)}$, as well as the means of the signal in both $M_D$ and $\Delta M$ are floated in the fit; all other parameters are fixed to the values obtained from fits to the corresponding MC sample. In each bin, the fit is performed simultaneously for $D^0$ and $\overline{D}^0$ categories to obtain the signal yield. Figure~\ref{Fig:Dst_total} shows the fit projections compared to the data within bin 1. These projections are signal-enhanced by considering events in the signal region of the variable that is not plotted; the signal regions are defined as $1.86 < M_D < 1.87$~GeV/$c^2$ and $0.144 < \Delta M < 0.146$~GeV/$c^2$. The large statistics of the sample makes it difficult for the model to fit data exactly, resulting in systematic deviations in the pull values from zero in the tails. Studies of MC samples have shown that the signal yield is unbiased and this systematic deviation in the pull values has negligible effect on the measured $K_i$ and $\overline{K}_i$ values. The efficiency- and migration-corrected yields are then used to determine the values of $K_i$ and $\overline{K}_i$, which are given in table~\ref{Table:Ki}. The values of $K_i$ and $\overline{K}_i$ are in reasonable agreement with those reported in ref.~\cite{Resmi}; the only deviation larger than $3\sigma$ is in bin 9, which contains only 1.2\% of the data.  

\begin{table} [t] 
\centering  
\begin{tabular} {l cc cc }
\hline 
Bin no. & $N_{D^0}$ & $N_{\overline{D}^0}$ & $K_i$ & $\overline{K}_i$ \\[0.5ex]
\hline
\hline
1& $\phantom{0}$51048$\pm$282 & $\phantom{0}$50254$\pm$280 & 0.2229$\pm$0.0008 & 0.2249$\pm$0.0008\\[0.5ex]
2& 137245$\pm$535 & $\phantom{0}$58222$\pm$382 & 0.4410$\pm$0.0009 & 0.1871$\pm$0.0007\\[0.5ex]
3& $\phantom{0}$31027$\pm$297 & 105147$\pm$476 & 0.0954$\pm$0.0005 & 0.3481$\pm$0.0009\\[0.5ex]
4& $\phantom{0}$24203$\pm$280 & $\phantom{0}$16718$\pm$246 & 0.0726$\pm$0.0005 & 0.0478$\pm$0.0004\\[0.5ex]
5& $\phantom{0}$13517$\pm$220 & $\phantom{0}$20023$\pm$255 & 0.0371$\pm$0.0003 & 0.0611$\pm$0.0004\\[0.5ex]
6& $\phantom{0}$21278$\pm$269 & $\phantom{0}$20721$\pm$267 & 0.0672$\pm$0.0005 & 0.0679$\pm$0.0005\\[0.5ex]
7& $\phantom{0}$15784$\pm$221 & $\phantom{0}$13839$\pm$209 & 0.0403$\pm$0.0004 & 0.0394$\pm$0.0004\\[0.5ex]
8& $\phantom{00}$6270$\pm$148 & $\phantom{00}$7744$\pm$164 & 0.0165$\pm$0.0002 & 0.0183$\pm$0.0002\\[0.5ex]
9& $\phantom{00}$6849$\pm$193 & $\phantom{00}$6698$\pm$192 & 0.0070$\pm$0.0002 & 0.0054$\pm$0.0001\\[0.5ex]

\hline
\end{tabular}
\caption{$D^0$ and $\overline{D}^0$ yield in each bin of $D$ phase space along with $K_i$ and $\overline{K}_i$ values measured in $D^{*}$ tagged data sample.}\label{Table:Ki}
\end{table}

\section{Determination of $(x_{\pm},y_{\pm})$ from the $B^{\pm}\to Dh^{\pm}$ sample}
\label{Sec:BtoDh}

We select both $B^{+}\to DK^{+}$ and $B^{+}\to D\pi^{+}$ decays because they have an identical topology, but the latter is less sensitive to $CP$-violation measurements because $r_{B}^{D\pi}$ is approximately twenty times smaller than $r_B^{DK}$. However, the $B^{+}\to D\pi^{+}$  branching fraction is an order of magnitude larger than that of $B^{+}\to DK^{+}$ and hence serves as an excellent calibration sample for the signal determination procedure. Furthermore, there is a significant background from $B^{+}\to D\pi^{+}$ decays in the $B^{+}\to DK^{+}$ sample from the misidentification of the charged pion as a kaon; a simultaneous fit to both samples allows this cross-feed to be directly determined from data. 

The signal yield in each bin is obtained via a simultaneous two-dimensional fit to the nine $D$ phase space bins with the data divided into $B^{+}\to DK^+$, $B^{-}\to DK^-$, $B^+\to D\pi^+$ and $B^-\to D\pi^-$ candidates, so there are 36 samples in total. The signal extraction is done by fitting $\Delta E$ and $C_{\rm NN}$. The distribution of $C_{\rm NN}$ cannot be described readily by an analytic PDF. Therefore, we transform $C_{\rm NN}$ as
\begin{equation}
C_{\rm NN}' = \log\left( \frac{C_{\rm NN} - C_{\rm NN,\rm low}}{C_{\rm NN,\rm high} - C_{\rm NN}} \right),
\end{equation}
where $C_{\rm NN,\rm low}$ = $-0.6$ and $C_{\rm NN,\rm high}$ = $0.9985$ are the minimum and maximum values of $C_{\rm NN}$ in the sample, respectively. The signal and background distributions of $C_{\rm NN}^{\prime}$ can be described by combinations of Gaussian PDFs. The three background components considered are:
\begin{itemize}
\item {\it continuum} background from $e^+e^- \to q\overline{q}$ processes, where $q = (u, d, s, c)$
\item {\it combinatorial  $B\overline{B}$} background, in which the final state particles could be coming from both $B$ mesons in an event; and
\item {\it cross-feed peaking background from $B^+ \to Dh^+$}, where $h=\pi,~K$, in which the charged kaon is misidentified as a pion or {\it vice versa}.
\end{itemize}

There is no significant correlation between $\Delta E$ and $C_{\rm NN}'$, so the two-dimensional PDF for each of the components is the product of one-dimensional $\Delta E$ and $C_{\rm NN}'$ PDFs.  The sum of a CB function and two Gaussian functions with a common mean is used as the PDF to model the $\Delta E$ signal component in both $B$ samples. The sum of a Gaussian and an asymmetric Gaussian with different mean values is used to parametrize the PDF that describes the $C_{\rm NN}'$ signal component. The continuum background distribution is modeled with a first-order polynomial in $\Delta E$ and by the sum of two Gaussian PDFs with different mean values in $C_{\rm NN}'$. The $\Delta E$ distribution of combinatorial $B\overline{B}$ background in $B^{+} \to D\pi^{+}$ is described by an exponential function. There is a small peaking structure due to misreconstructed $\pi^0$ events, and this is modeled by a CB function. A first-order polynomial is added to the above two PDFs in the case of $B^{+}\to DK^{+}$ decays. The $C_{\rm NN}'$ distribution for each of the samples is modeled by an asymmetric Gaussian function. The cross-feed peaking background in $\Delta E$ is modeled with the sum of three Gaussian functions, whereas the signal PDF itself is used for the $C_{\rm NN}'$ distribution. 

\begin{figure}[t]
\centering
\includegraphics[width=\columnwidth]{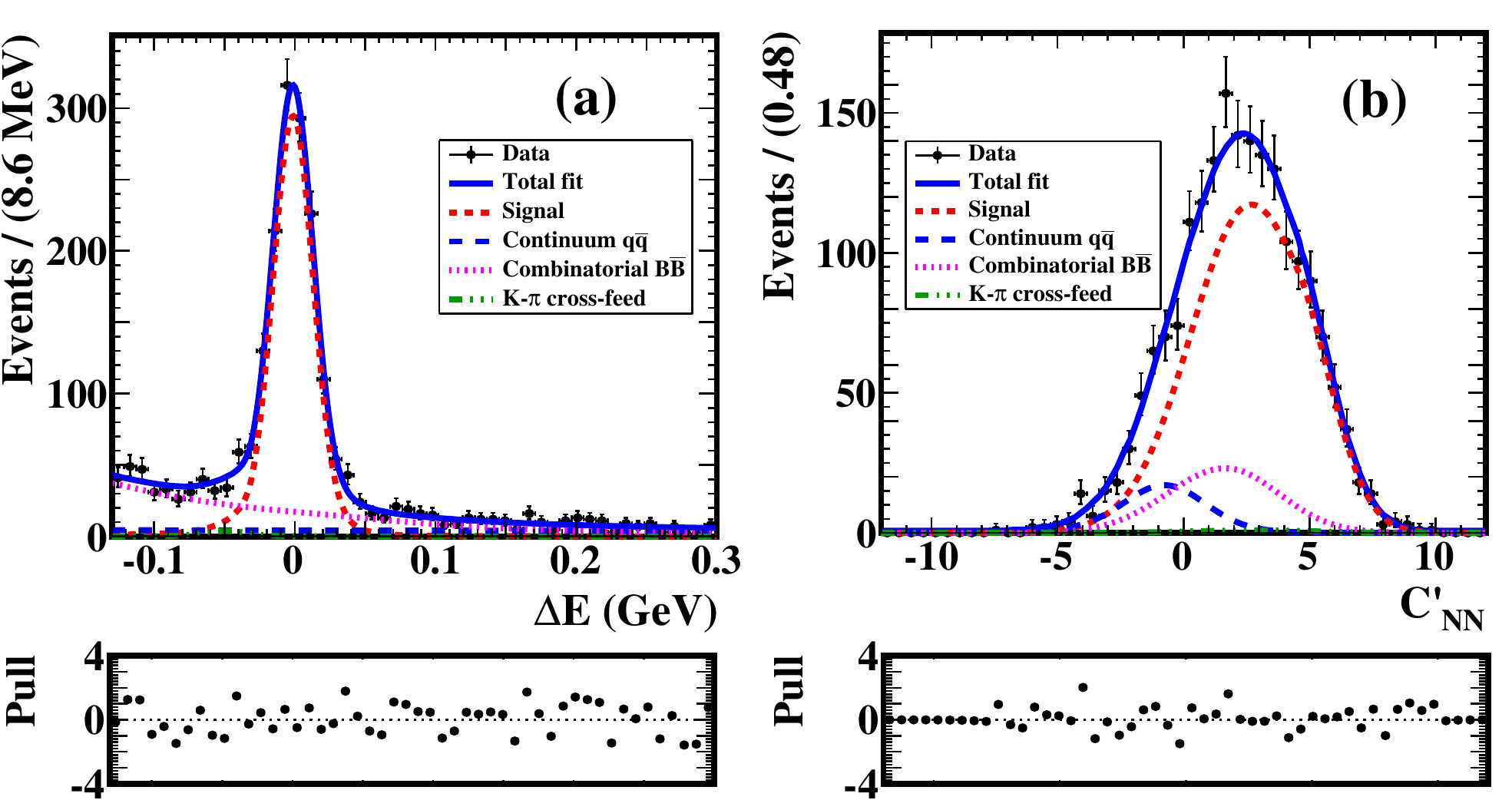}
\caption{Signal-enhanced fit projections of (a) $\Delta E$ and (b) $C_{\rm NN}'$ for the $B^{\pm}\to D\pi^{\pm}$ data sample in bin 1. The black points with error bars are the data and the solid blue curves are the total fit. The dotted red, blue, magenta, and green curves represent the signal, continuum, combinatorial $B\overline{B}$ backgrounds and cross-feed peaking background components, respectively. The pull between the data and the fit is shown for both the projections.}\label{Fig:dpi}
\end{figure}

\begin{figure}[t]
\centering
\includegraphics[width=\columnwidth]{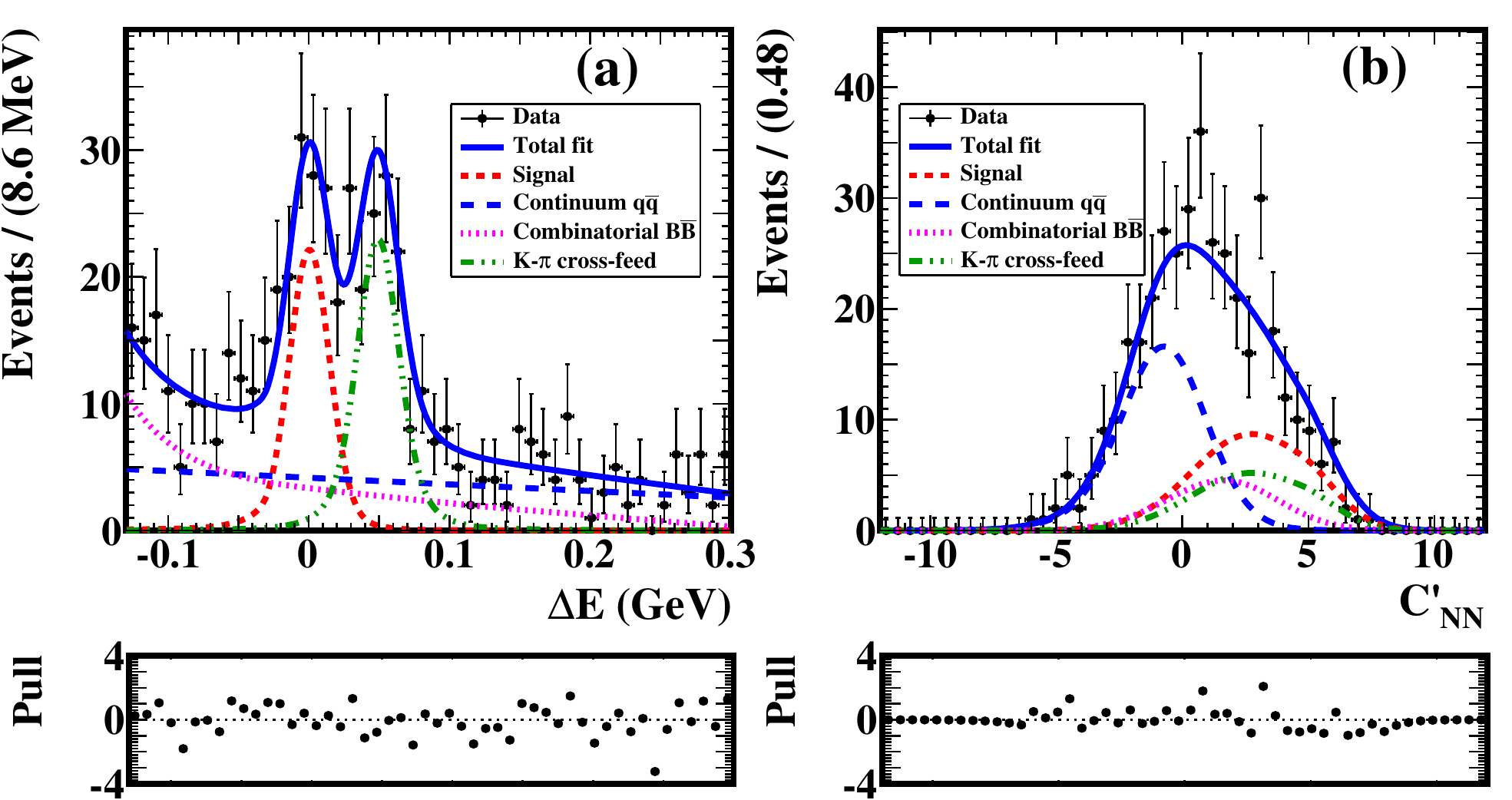}
\caption{Signal-enhanced fit projections of (a) $\Delta E$ and (b) $C_{\rm NN}'$ for the $B^{\pm}\to DK^{\pm}$ data sample in bin 1. The black points with error bars are the data and the solid blue curves are the total fit. The dotted red, blue, magenta, and green curves represent the signal, continuum, combinatorial $B\overline{B}$ backgrounds and cross-feed peaking background components, respectively. The pull between the data and the fit is also shown for both the projections. }\label{Fig:dk}
\end{figure}

 All yields are determined from the fit to data. The signal mean value and polynomial parameter for continuum background $\Delta E$ distribution are determined from the fit to data, while all other shape parameters are fixed to those obtained from fits to appropriate MC samples. A scaling factor is applied on the $\Delta E$ signal resolution, which is a free parameter in the fit. All $C_{\rm NN}'$ parameters are fixed to the values obtained from MC. An additional shift is applied on the continuum background mean value as well as a scaling factor to the resolution. Both these parameters are determined from data, which ensures that any possible data-MC difference is taken into account. We do not perform an independent fit in each bin because the event yields become too small to determine all the free parameters. Therefore, common shape parameters are used for each bin except for the combinatorial $B\overline{B}$ background component in bin 1. A separate exponential parameter is used in bin 1 due to the difference in slope compared to other bins. These exponential parameters are also floated in the fit in addition to those mentioned earlier. The signal-enhanced fit projections for the data in bin 1 are shown in figure~\ref{Fig:dpi} and \ref{Fig:dk}, where the signal regions are defined as $|\Delta E| < 0.05$ GeV and $ C_{\rm NN}' >0$.   The fitted signal yields are summarized in table~\ref{Table:dpidk}. The total numbers of $B^{\pm}\to D\pi^{\pm}$  and  $B^{\pm}\to DK^{\pm}$ signal events are 9981 $\pm$ 134 and 815 $\pm$ 51, respectively.

\begin{table} [t] 
\centering 
\begin{tabular} {l c c c c  }
\hline 
 & \multicolumn{2}{c}{$B^{\pm}\to D\pi^{\pm}$} & \multicolumn{2}{c}{$B^{\pm}\to DK^{\pm}$}\\[0.5ex]
\hline
Bin no. & $N_i^+$ & $N_i^-$ & $N_i^+$ & $N_i^-$\\[0.5ex]
\hline
\hline
1  & $\phantom{1}$772$\pm$33 & $\phantom{2}$860$\pm$34 & $\phantom{1}$80$\pm$13 & $\phantom{0}$58$\pm$12\\[0.5ex]
2  & 1077$\pm$41 & 2088$\pm$55 & $\phantom{1}$98$\pm$16 & 190$\pm$21\\[0.5ex]
3  & 1639$\pm$49 & $\phantom{2}$450$\pm$28& 121$\pm$18 & $\phantom{0}$57$\pm$13 \\[0.5ex]
4 & $\phantom{1}$263$\pm$24 & $\phantom{2}$451$\pm$29 & $\phantom{1}$21$\pm$9$\phantom{0}$ & $\phantom{0}$30$\pm$11\\[0.5ex]
5  &$\phantom{1}$377$\pm$27 & $\phantom{2}$256$\pm$23 & $\phantom{1}$23$\pm$9$\phantom{0}$ & $\phantom{0}$18$\pm$9$\phantom{0}$\\[0.5ex]
6  &$\phantom{1}$338$\pm$26 & $\phantom{2}$321$\pm$26 & $\phantom{1}$35$\pm$11 & $\phantom{0}$23$\pm$9$\phantom{0}$\\[0.5ex]
7  & $\phantom{1}$253$\pm$21 & $\phantom{2}$255$\pm$22 & $\phantom{1}$16$\pm$9$\phantom{0}$ & $\phantom{01}$5$\pm$7$\phantom{0}$\\[0.5ex]
8  &$\phantom{1}$154$\pm$17 & $\phantom{2}$109$\pm$15& $\phantom{10}$9$\pm$6$\phantom{0}$ & $\phantom{0}$13$\pm$7$\phantom{0}$ \\[0.5ex]
9  &$\phantom{1}$162$\pm$19 & $\phantom{2}$138$\pm$19 & $\phantom{1}$21$\pm$9$\phantom{0}$ & $\phantom{1}$30$\pm$10\\[0.5ex]
\hline
\end{tabular}
\caption{Signal yields in each $D$ phase space bin for $B^{\pm}\to D\pi^{\pm}$ and $B^{\pm}\to DK^{\pm}$ data samples obtained from a simultaneous fit to the nine bins.}\label{Table:dpidk}

\end{table}

\begin{figure}[t]
\centering
\begin{tabular}{cc}
\includegraphics[width=0.5\columnwidth]{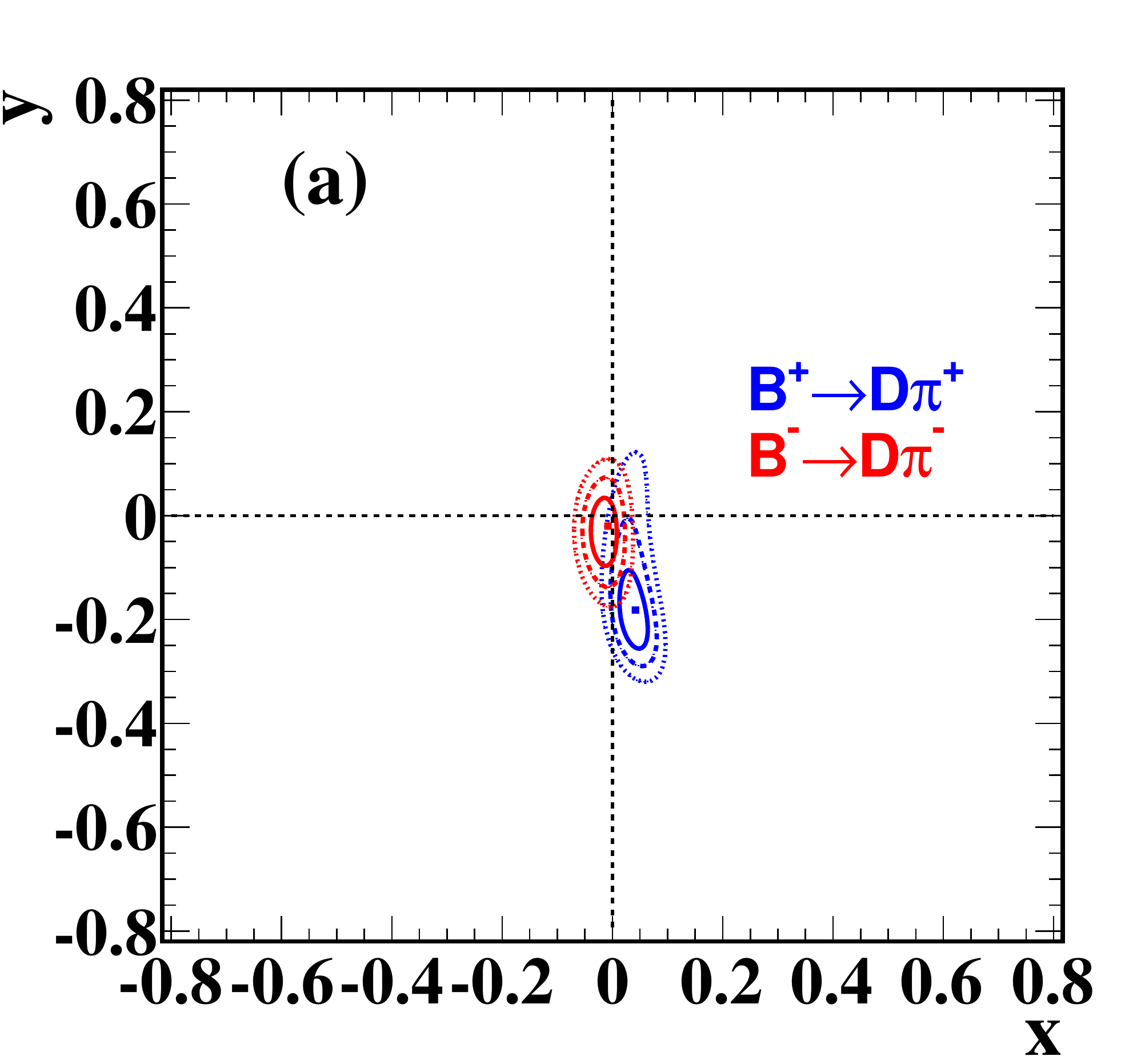}&
\includegraphics[width=0.5\columnwidth]{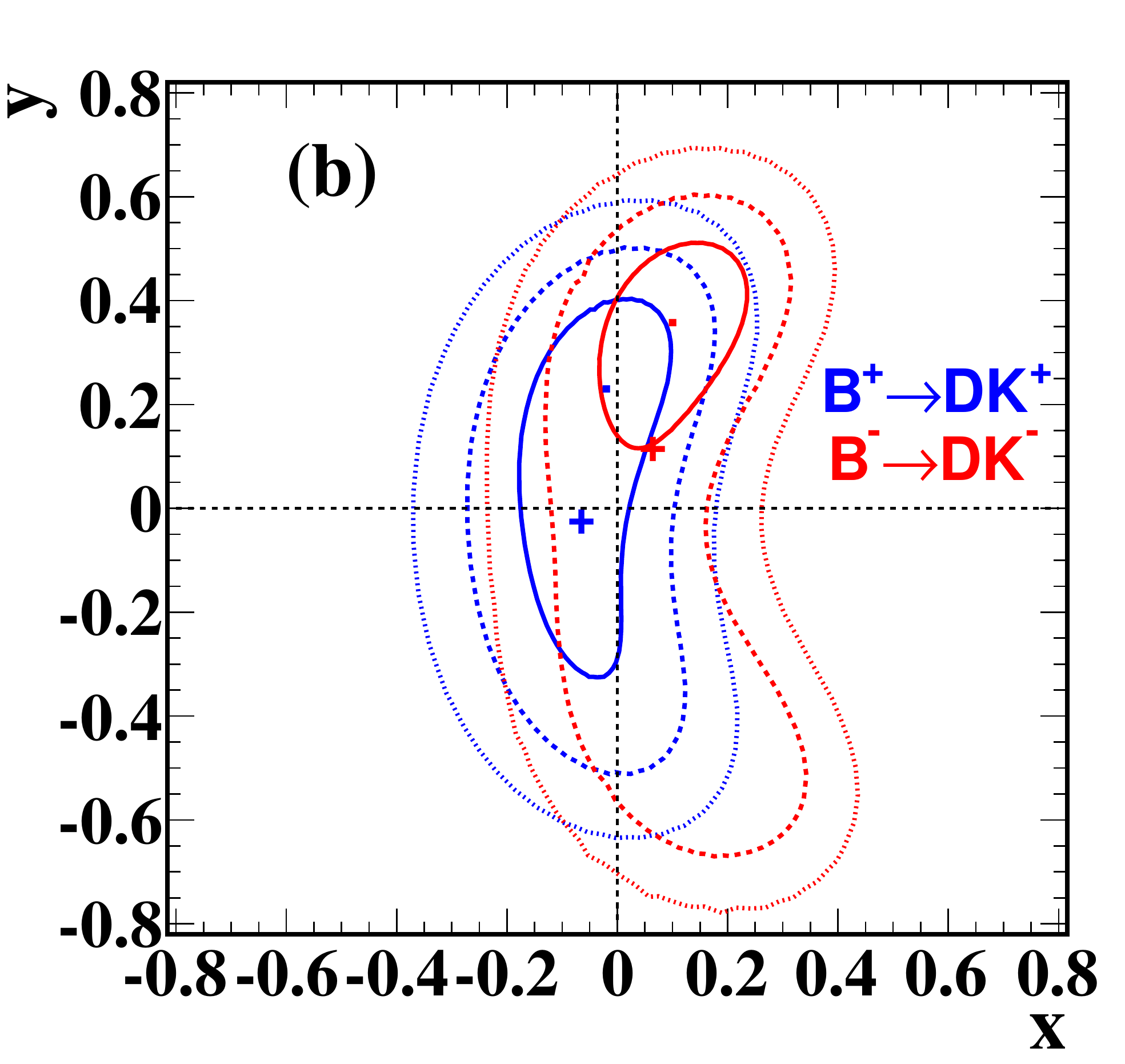}\\
\end{tabular}
\caption{One (solid line), two (dashed line), and three (dotted line) standard deviation  likelihood contours for the $(x_{\pm},y_{\pm})$ parameters for (a) $B^{\pm}\to D\pi^{\pm}$ and (b) $B^{\pm}\to DK^{\pm}$ decays. The point marks the best fit value and the cross marks the expected value from the world average values of $\phi_3$, $r_{B}^{DK}$, and $\delta_{B}^{DK}$ \cite{HFLAV}.}
\label{Fig:contour}
\end{figure}

\begin{figure}[t]
\centering
\begin{tabular}{cc}
\includegraphics[width=0.5\columnwidth]{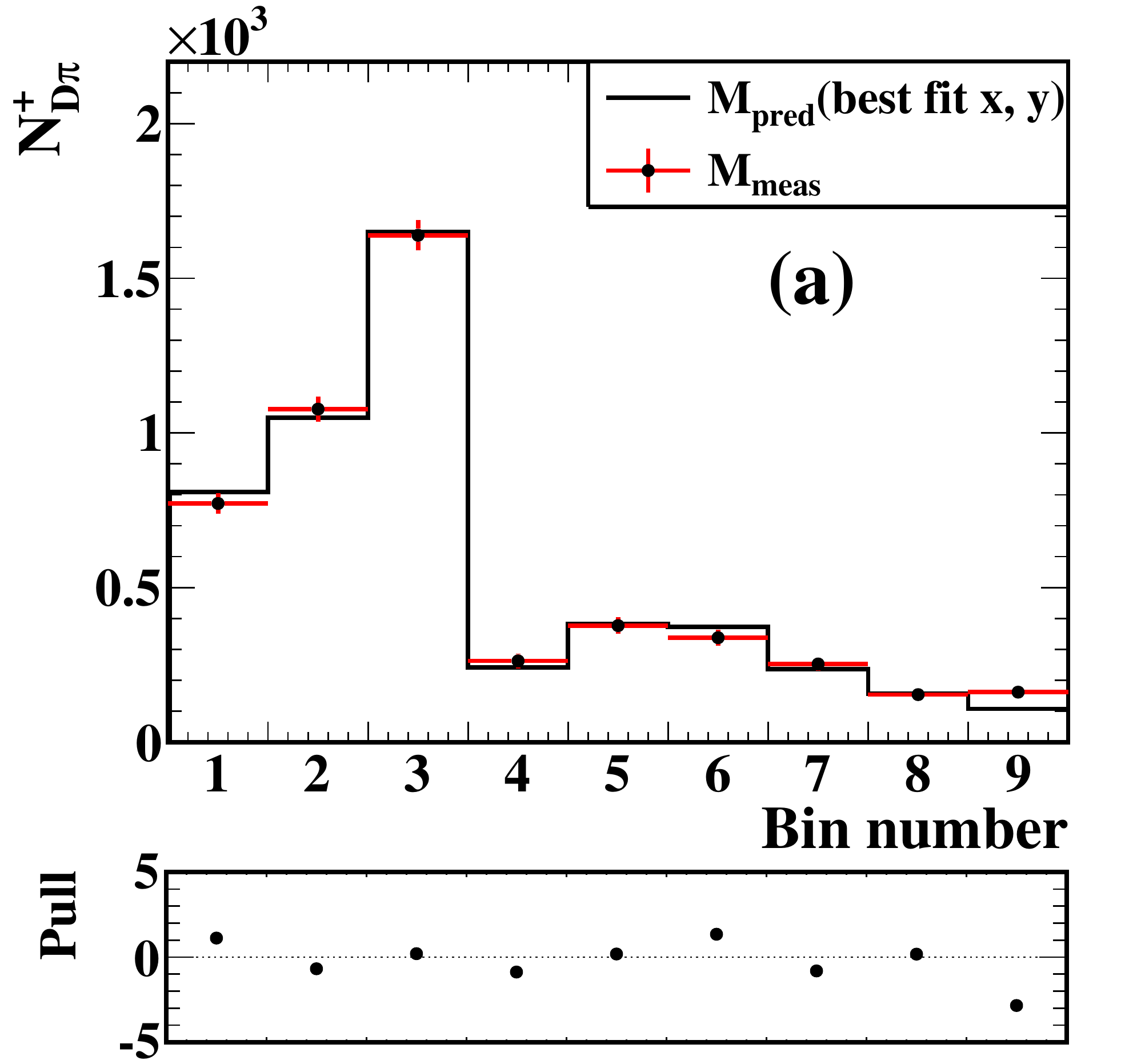}&
\includegraphics[width=0.5\columnwidth]{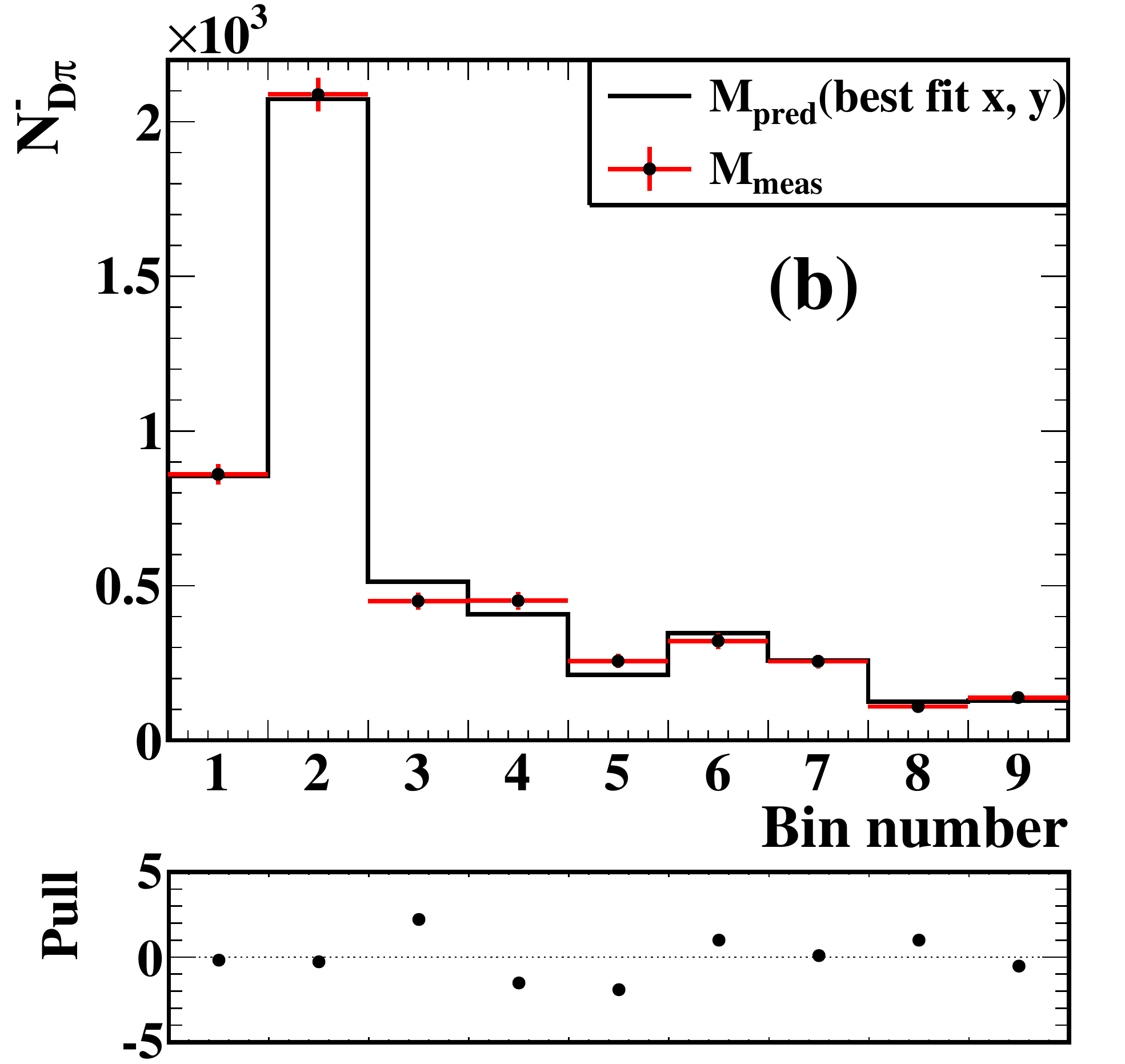}\\
\end{tabular}
\caption{Measured and expected yields in bins for (a) $B^+\to D\pi^+$ and (b) $B^-\to D\pi^-$ data samples. The data points with error bars are the measured yields, and the solid histogram is the expected yield from the best fit  $(x_{\pm},y_{\pm})$  parameter values.}
\label{Fig:dpi_exp}
\end{figure}

\begin{figure}[t]
\centering
\begin{tabular}{cc}
\includegraphics[width=0.5\columnwidth]{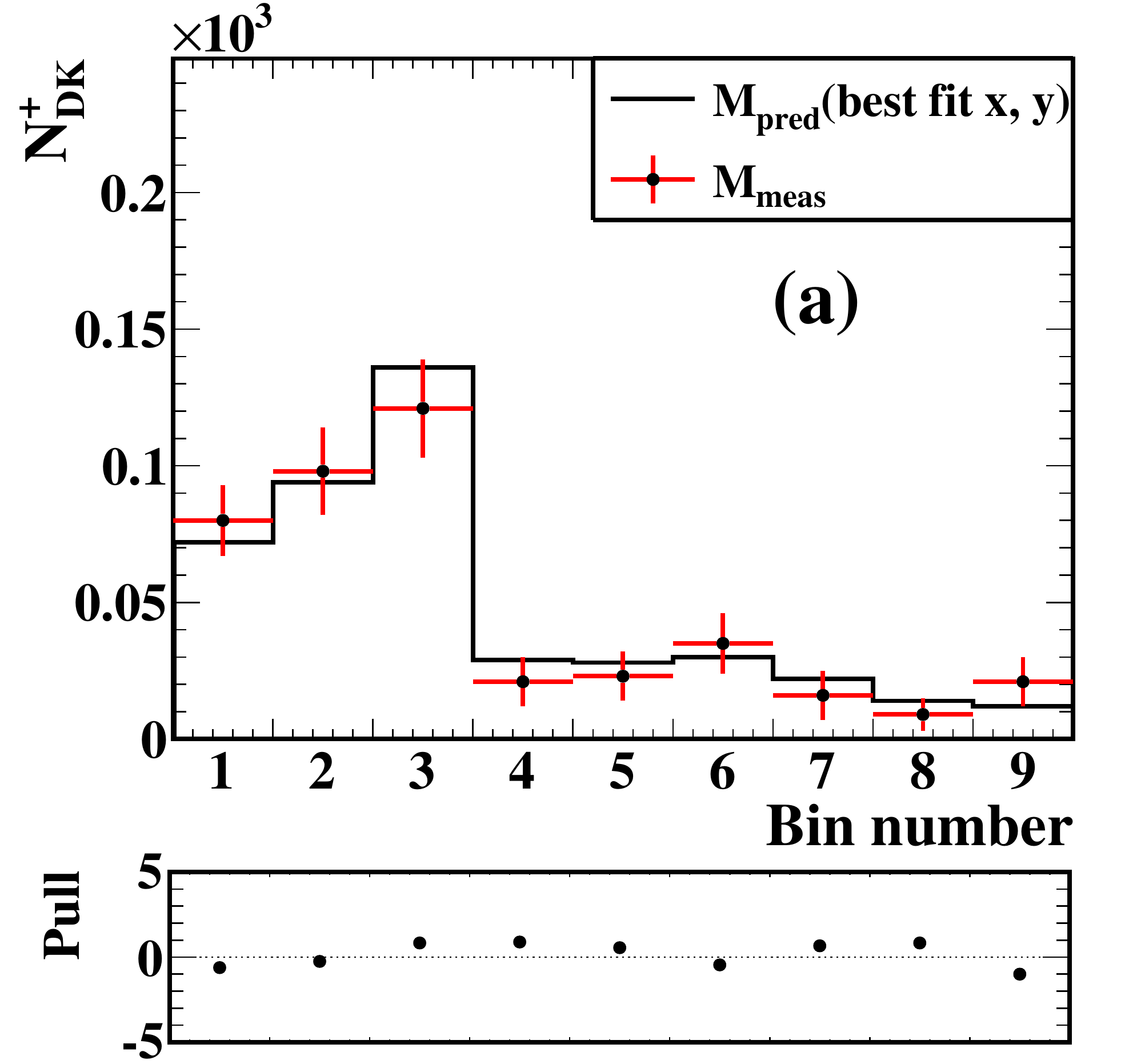}&
\includegraphics[width=0.5\columnwidth]{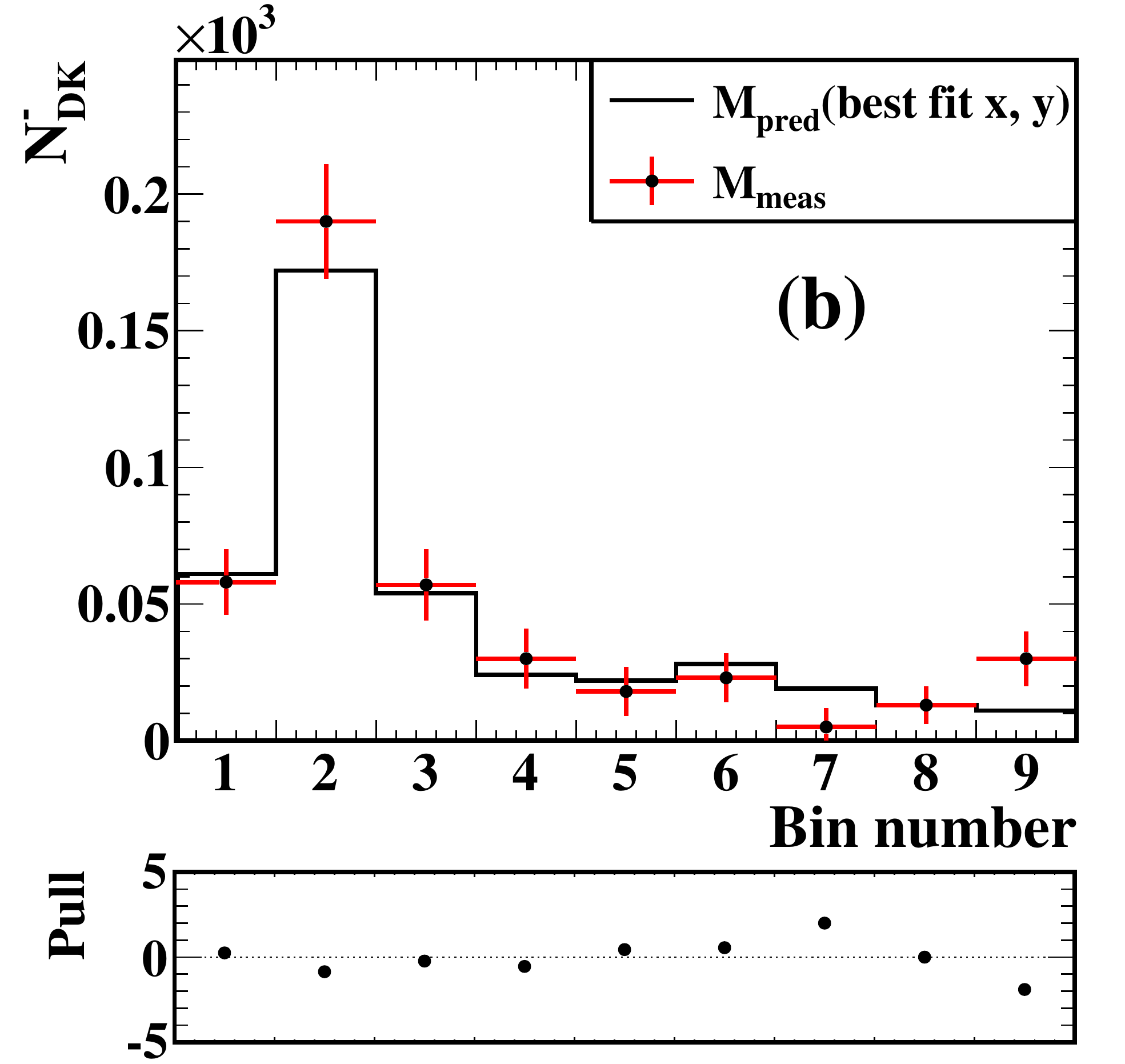}\\
\end{tabular}
\caption{Measured and expected yields in bins for (a) $B^+\to DK^+$ and (b) $B^-\to DK^-$ data samples. The data points with error bars are the measured yields, and the solid histogram is the expected yield from the best fit  $(x_{\pm},y_{\pm})$  parameter values.}
\label{Fig:dk_exp}
\end{figure}

The Cartesian parameters $x_{\pm}$ and $y_{\pm}$ are extracted from the simultaneous fit by expressing the signal yield using eqs.~\eqref{Eq:B-} and \eqref{Eq:B+}; the procedure includes corrections for efficiency and migration between bins. The input parameters to the expressions in  eqs.~\eqref{Eq:B-} and \eqref{Eq:B+}~include the values of $K_i$ and $\overline{K}_i$ obtained from the flavour-tagged $D$ sample and the $D$ strong-phase difference parameters $c_i$ and $s_i$~\cite{Resmi}. The results are summarized in table~\ref{Table:xy}, and the statistical likelihood contours are shown in figure~\ref{Fig:contour}. The statistical correlation matrices are given in tables~\ref{Table:xycorrdpi} and \ref{Table:xycorr}. The measured and expected yields  for the binned $B^+$ and $B^-$ data are compared in figures~\ref{Fig:dpi_exp} and \ref{Fig:dk_exp}. 
\begin{table} [t] 
\centering  
\begin{tabular} {c c c }
\hline 
& $B^{\pm}\to D\pi^{\pm}$ & $B^{\pm}\to DK^{\pm}$ \\[1ex]
\hline
\hline
$x_{+}$ & 0.039 $\pm$ 0.024~$^{+0.018~+0.014}_{-0.013~-0.012}$ & $-$0.030 $\pm$ 0.121~$^{+0.017~+0.019}_{-0.018~-0.018}$     \\[1ex]

$y_{+}$ & $-$0.196~$^{+0.080~+0.038~+0.032}_{-0.059~-0.034~-0.030}$ & 0.220~$^{+0.182}_{-0.541}$  $\pm$ 0.032~$^{+0.072}_{-0.071}$   \\[1ex]

$x_{-}$ & $-$0.014 $\pm$0.021~$^{+0.018~+0.019}_{-0.010~-0.010}$ & 0.095 $\pm$ 0.121~$^{+0.017~+0.023}_{-0.016~-0.025}$ \\[1ex]

$y_{-}$  & $-$0.033 $\pm$~0.059$^{+0.018~+0.019}_{-0.019~-0.010}$ & 0.354~$^{+0.144~+0.015~+0.032}_{-0.197~-0.021~-0.049}$ \\[1ex]
\hline

\end{tabular}
\caption{$x_{\pm}$and $y_{\pm}$ parameters from a combined fit to $B^{\pm}\to D\pi^{\pm}$ and $B^{\pm}\to DK^{\pm}$ data samples. The first uncertainty is statistical, the second is systematic, and the third is due to the uncertainty on the $c_i$, $s_i$ measurements. }\label{Table:xy}
\end{table}
 \begin{table} [t] 
\centering  
\begin{tabular} {c c c c  c  }
\hline 
& $x_{+}$ & $y_{+}$ & $x_{-}$ & $y_{-}$ \\[1ex]
\hline
\hline
$x_{+}$& 1 & $-$0.364 & 0.314 &  $\phantom{-}$0.050 \\[0.5ex]
 $y_{+}$& & 1& 0.347 &  $\phantom{-}$0.055 \\[0.5ex]
 $x_{-}$ & & & 1 & $-$0.032\\[0.5ex]
 $y_{-}$ & & & & 1\\[0.5ex]
\hline

\end{tabular}
\caption{Statistical correlation matrix for $(x_{+}, y_{+}, x_{-}, y_{-})$ measured from the $B^{\pm}\to D\pi^{\pm}$ data sample}\label{Table:xycorrdpi}
\end{table}
\begin{table} [t] 
\centering  
\begin{tabular} {c c c c  c  }
\hline 
& $x_{+}$ & $y_{+}$ & $x_{-}$ & $y_{-}$ \\[1ex]
\hline
\hline
$x_{+}$& 1 & 0.486 &  $\phantom{-}$0.172 & $-$0.231 \\[0.5ex]
 $y_{+}$& & 1& $-$0.127 & $\phantom{-}$0.179 \\[0.5ex]
 $x_{-}$ & & & 1 &  $\phantom{-}$0.365\\[0.5ex]
 $y_{-}$ & & & & 1\\[0.5ex]
\hline

\end{tabular}
\caption{Statistical correlation matrix for $(x_{+}, y_{+}, x_{-}, y_{-})$ measured from the $B^{\pm}\to DK^{\pm}$ data sample}\label{Table:xycorr}
\end{table}

\section{Systematic uncertainties}
\label{Sec:syst}

We consider several possible sources of systematic uncertainty, as listed in table~\ref{Table:syst}, along with their contributions. The remainder of this section describes how these uncertainties are estimated.

The limited size of the signal MC sample used for estimating the efficiency and the migration matrix is a source of systematic uncertainty. Efficiencies in $B$ and $D^*$ samples are varied by their statistical uncertainty ($\pm 1\sigma$) in each bin independently. The resultant negative and positive deviations in  $(x_{\pm},y_{\pm})$  are separately summed in quadrature. Similarly, the migration matrix elements are varied by their statistical uncertainty in $B$ and $D^*$ samples, one element at a time. The resultant positive and negative deviations are considered separately.

The systematic uncertainty due to the difference in mass resolution between data and the MC samples is considered by varying the width on the $\pi^+\pi^-\pi^0$ invariant mass distribution by the uncertainty on the resolution scale factor obtained in data, when compared to that in MC. The resultant deviations in  $(x_{\pm},y_{\pm})$  are taken as the systematic uncertainty from this source. All the other resonances are wide and the resolution difference is an order of magnitude smaller than the resolution, thus the modelling of resolution does not affect our measurements. The systematic effect of the uncertainty on the $K_i$ and $\overline{K}_i$ values is estimated by varying them by their statistical uncertainties independently. The resultant sum of deviations in quadrature is taken as the associated systematic uncertainty. 

Modelling the data with PDFs that have parameters fixed to values obtained from MC samples is another  source of systematic uncertainty. There are 14 signal and 23 background shape parameters fixed in the $B^{\pm}\to Dh^{\pm}$ simultaneous fit. These are fixed to the values obtained from MC samples. The uncertainty due to PDF modelling is taken into account by repeating the fit by individually varying the fixed parameters by $\pm 1\sigma$, where $\sigma$ is the uncertainty on these parameters in MC component fits, and taking the difference in quadrature as the uncertainty. Any possible bias in the fit is studied with a set of pseudo-experiments with different input values for  $(x_{\pm},y_{\pm})$. The fit is found to give an unbiased response within the statistical uncertainty from the finite number of pseudo-experiments, and this uncertainty is taken as the systematic uncertainty from this source.

The kaon identification efficiency and pion fake rate used in the fit are also fixed parameters that are determined from control samples of $D^{*+}\to D^{0}\pi^{+}$, $D^{0}\to K^-\pi^+$. They are varied by $\pm 1 \sigma$ and the resultant deviations in the nominal  $(x_{\pm},y_{\pm})$  values are assigned as the systematic uncertainty. The uncertainty on the $c_i,~ s_i$ inputs reported in ref.~\cite{Resmi} are also considered by varying  $c_i,~ s_i$  by their respective uncertainties and then considering the corresponding deviations in $(x_{\pm},y_{\pm})$ from the nominal values as the systematic uncertainty. Here, the correlation between  $c_i,~ s_i$ is taken into account. The effect of the difference in the efficiency variation across the bins for $B$ and $D^*$ samples is studied. We find no deviation in $K_i$ and $\overline{K}_i$ values within their statistical uncertainty when the $D^*$ efficiencies are varied by the maximum deviation found between the samples or $D$ momentum range is changed to 1\textendash 3 GeV/$c$. 
\begin{table} [t] 
\centering
 \footnotesize  
 \begin{tabular} {c cccc cccc }
\hline
Source& \multicolumn{4}{c}{ $B^{\pm}\to D\pi^{\pm}$} & \multicolumn{4}{c}{ $B^{\pm}\to DK^{\pm}$}\\[0.5ex]
\hline
& $x_{+}$ & $y_{+}$ & $x_{-}$ & $y_{-}$ & $x_{+}$ & $y_{+}$ & $x_{-}$ & $y_{-}$ \\[0.5ex]
\hline
\hline

Efficiency  & +0.013 & +0.030 & +0.012& +0.012 & +0.012 & +0.022 & +0.012& +0.013\\[0.5ex]
uncertainty& $-$0.009 & $-$0.027 & $-$0.008 &$-$0.013& $-$0.013 & $-$0.023 & $-$0.012 &$-$0.016 \\[1ex] \hline
Migration matrix  & +0.011 & +0.021 & +0.011& +0.013& +0.007 & +0.015 & +0.007& +0.006 \\[0.5ex] \
uncertainty& $-$0.004 & $-$0.019 & $-$0.003 &$-$0.014& $-$0.008 & $-$0.016 & $-$0.007 &$-$0.012 \\[1ex]
\hline
$m_{\pi\pi\pi^0}$ resolution & $\phantom{-}$0.003 & $\phantom{-}$ 0.001 &  $\phantom{-}$0.004 &  $\phantom{-}$0.001&  $\phantom{-}$0.001 &  $\phantom{-}$0.001 &   $\phantom{-}$0.001 &  $\phantom{-}$0.003 \\[1ex]
\hline

$K_i$, $\overline{K}_i$ & +0.004 & +0.007 & +0.004& +0.002& +0.001 & +0.001 & +0.002& +0.001 \\[0.5ex]
uncertainty& $-$0.001 & $-$0.006 & $-$0.001 &$-$0.002& $-$0.002 & $-$0.001 & $-$0.002 &$-$0.001 \\[1ex]
\hline 
PDF shape  & +0.004 & +0.004 & +0.004& +0.001& +0.009 & +0.017 & +0.009& +0.001 \\[0.5ex]
 & $-$0.008 & $-$0.003 & $-$0.004 &$-$0.001& $-$0.008 & $-$0.016 & $-$0.007 &$-$0.005  \\[1ex]
\hline
Fit bias & $\phantom{-}$0.000 &  $\phantom{-}$0.001 &  $\phantom{-}$0.000 &  $\phantom{-}$0.000 &  $\phantom{-}$0.001 &  $\phantom{-}$0.001 &  $\phantom{-}$0.001 &  $\phantom{-}$0.003\\[1ex]
\hline 
PID &  $\phantom{-}$0.001 &  $\phantom{-}$0.001 &  $\phantom{-}$0.001 &  $\phantom{-}$0.000  &  $\phantom{-}$0.002 &  $\phantom{-}$0.001 &  $\phantom{-}$0.002 &  $\phantom{-}$0.001 \\[1ex]
\hline
\hline
Total systematic & +0.018 & +0.038 & +0.018& +0.018 & +0.017 & +0.032 & +0.017& +0.015 \\[0.5ex]
uncertainty & $-$0.013 & $-$0.034 & $-$0.010 &$-$0.019& $-$0.018 & $-$0.032 & $-$0.016 &$-$0.021 \\[1ex] 
 \hline
 \hline
$c_i, s_i$ & +0.014 & +0.032 & +0.010& +0.019 & +0.019 & +0.072 & +0.023& +0.032\\[0.5ex]
uncertainty & $-$0.012 & $-$0.030 & $-$0.006 &$-$0.010& $-$0.018 & $-$0.071 & $-$0.025 &$-$0.049 \\[1ex]
\hline
\hline
Total statistical & +0.024 & +0.080 & +0.021& +0.059 & +0.121 & +0.182 & +0.121& +0.144 \\[0.5ex]
uncertainty & $-$0.024 & $-$0.059 & $-$0.021 &$-$0.059& $-$0.121 & $-$0.541 & $-$0.121 &$-$0.197 \\[1ex] 
 \hline
\end{tabular}
\normalsize
\caption{Systematic uncertainties from various sources in $B^{\pm}\to D\pi^{\pm}$ and $B^{\pm}\to DK^{\pm}$ data samples.}\label{Table:syst}
\end{table}

\section{Determination of $\phi_3,~r_B$ and $\delta_B$}
\label{Sec:phi3}

We use the frequentist treatment, which includes the Feldman-Cousins ordering~\cite{FC}, to obtain the physical parameters $$\mu = (\phi_3, r_{B}, \delta_{B})\; ,$$ from the measured parameters $$z = (x_{+}, y_{+}, x_{-}, y_{-})\; , $$ in $B^{\pm}\to DK^{\pm}$ sample; this is the same procedure as was used in ref. \cite{Belle-GGSZ}. We do not use the $B^{\pm}\to D\pi^{\pm}$ sample to constrain $\phi_3$, which has been the case in previous Belle analyses \cite{Belle-GGSZ, Belle-GGSZ2}. We note that the constraints presented by the LHCb Collaboration~\cite{LHCbconf1} allow values up to $r_B^{D\pi} <0.028$ at a 2$\sigma$ confidence level, which is five times larger than the expectation; if the value of $r_B^{D\pi}$ is significantly larger than expected then future analyses could include $B^{\pm}\to D\pi^{\pm}$ channel to determine $\phi_3$.
 The confidence level is calculated as 
 \begin{equation}
 \alpha(\mu) = \frac{\int_{\mathcal{D}(\mu)} p(z|\mu) dz}{\int_{\infty} p(z|\mu)dz},
 \end{equation}
 where $p(z|\mu)$ is the probability density to observe the measurements $z$ given the set of physical parameters $\mu$. The integration domain $\mathcal{D}(\mu)$ is given by the likelihood ratio ordering in the Feldman-Cousins method. The PDF $p(z|\mu)$ is a multivariate Gaussian PDF with the uncertainties and correlations between  $(x_{\pm},y_{\pm})$  taken from the measurements.
 
\begin{figure}[ht!]
\centering
\begin{tabular}{cc}
\includegraphics[width=0.5\columnwidth]{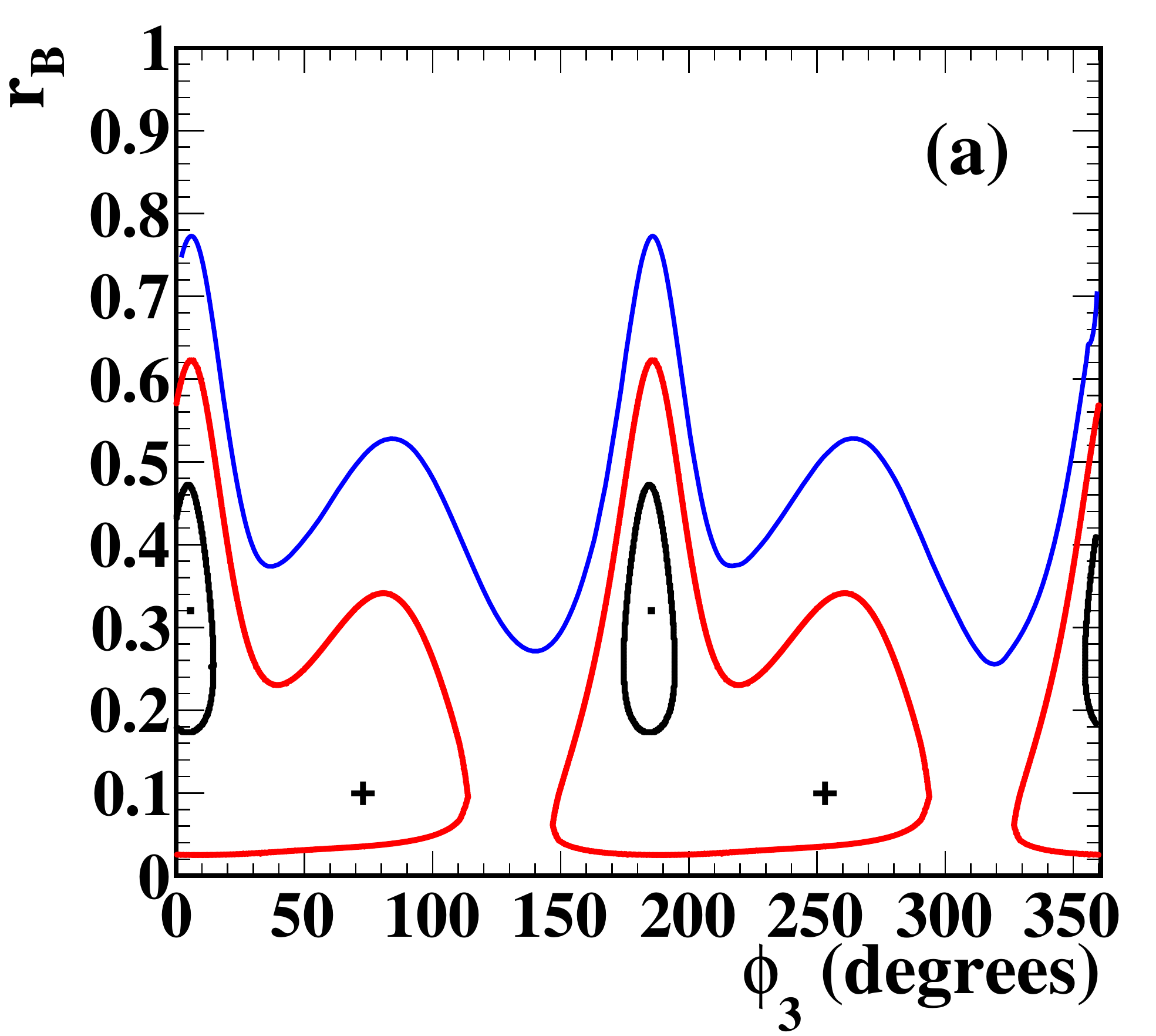}&
\includegraphics[width=0.5\columnwidth]{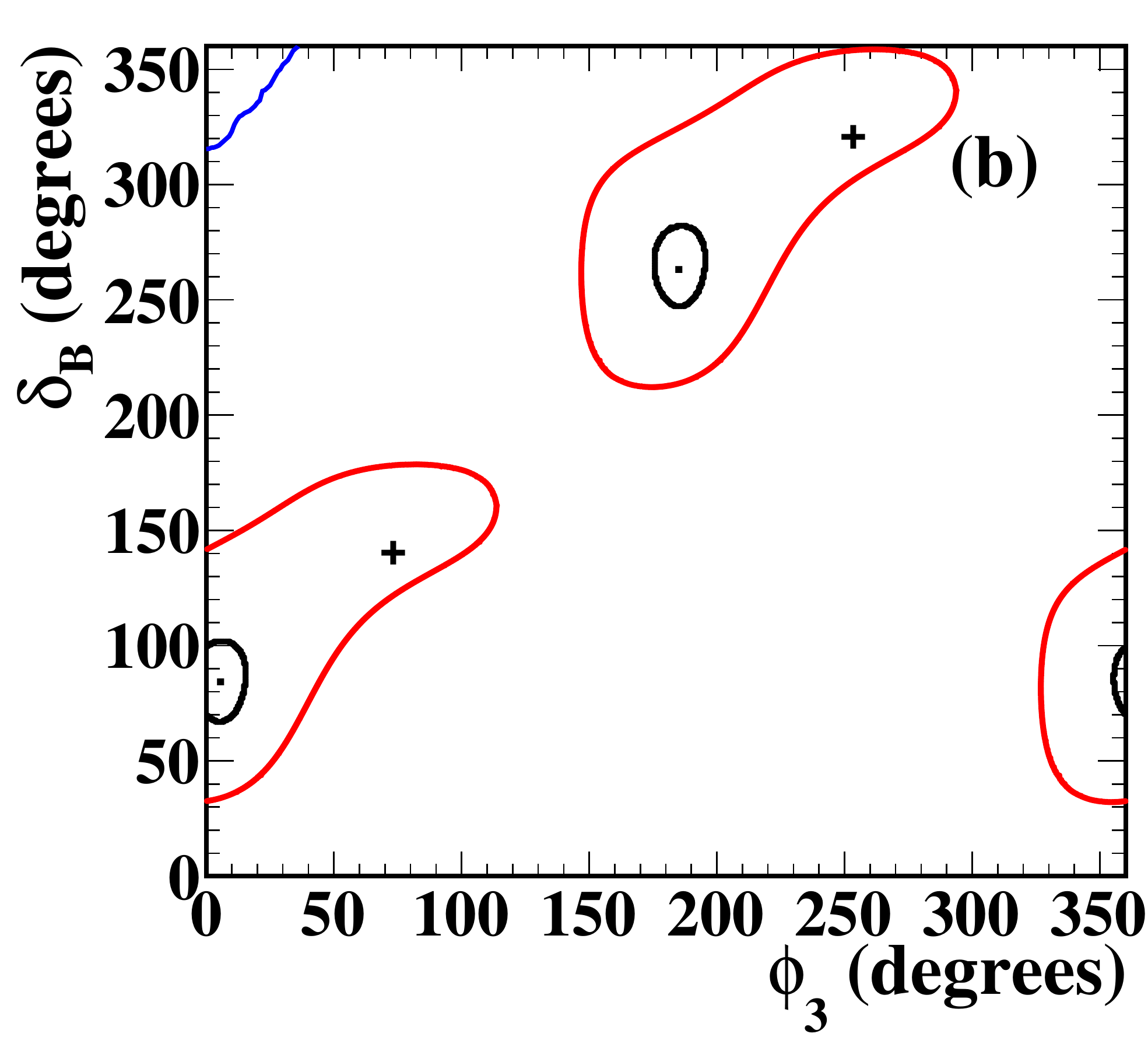}\\
\end{tabular}
\caption{Projection of the statistical confidence intervals in the (a) $\phi_3 - r_B$  and (b) $\phi_3 - \delta_B$  planes. The black, red, and blue contours represent the one, two, and three standard deviation regions, respectively. The crosses show the positions of the world-average values \cite{HFLAV}.}
\label{Fig:contours}
\end{figure}

 We obtain the parameters $\mu = (\phi_3, r_{B}, \delta_{B})$ from the fit as given in table~\ref{Table:phi3}. The systematic uncertainty is estimated by varying the $z$ parameters by their corresponding systematic uncertainties. Figure~\ref{Fig:contours} shows the confidence level contours representing one, two, and three standard deviations in $(\phi_3, r_B)$ and $(\phi_3, \delta_B)$ planes. 
\begin{table} [t] 
\centering  
\begin{tabular} {c c c c  c  }
\hline 
Parameter & Results & 2$\sigma$ interval\\[1ex]
\hline
\hline
$\phi_3$ ($^{\circ}$) & $ 5.7~^{+10.2}_{-8.8}~ \pm~ 3.5~ \pm~ 5.7$  & ($-$29.7, 109.5)\\[0.5ex]
$\delta_B$ ($^{\circ}$) & $83.4~^{+18.3}_{-16.6}~ \pm~ 3.1~ \pm~ 4.0 $ & (35.7, 175.0)\\[0.5ex]
$r_B$ & $0.323~ \pm~ 0.147~ \pm~ 0.023~ \pm~ 0.051$& (0.031, 0.616)\\[0.5ex]
\hline

\end{tabular}
\caption{$(\phi_3, \delta_B , r_B)$ obtained from the $B^{\pm}\to DK^{\pm}$ data sample. The first uncertainty is statistical, second is systematic and, the third one is due to the uncertainty on $c_i$, $s_i$ measurements.}\label{Table:phi3}
\end{table}

We performed a check of the assumption that the $(x_{\pm},y_{\pm})$ likelihood can be approximated to be Gaussian when using the Feldman-Cousins method to extract $(\phi_3,r_B,\delta_B)$. The check used the measured confidence intervals in $(\phi_3,r_B,\delta_B)$ to generate an ensemble of simulated data sets. Each simulated data set was then fit to form a  distribution of $(x_{\pm},y_{\pm})$, which was found to be consistent with the $(x_{\pm},y_{\pm})$ confidence intervals measured. Hence we conclude that the reported confidence intervals for $(\phi_3,r_B,\delta_B)$ are appropriate.

There is a two-fold ambiguity in $\phi_3$ and $\delta_B$ results with $\phi_3 + 180^{\circ}$ and $\delta_B + 180^{\circ}$. We choose the solution that satisfies $0^{\circ} < \phi_3 < 180^{\circ}$. This result includes the current world-average value~\cite{HFLAV} within two standard deviations. We observe that there is a local minimum of the likelihood around $\phi_3 = 75^{\circ}$ and $\delta_B = 155^{\circ}$.

\begin{figure}[t]
\centering
\includegraphics[width=0.5\columnwidth]{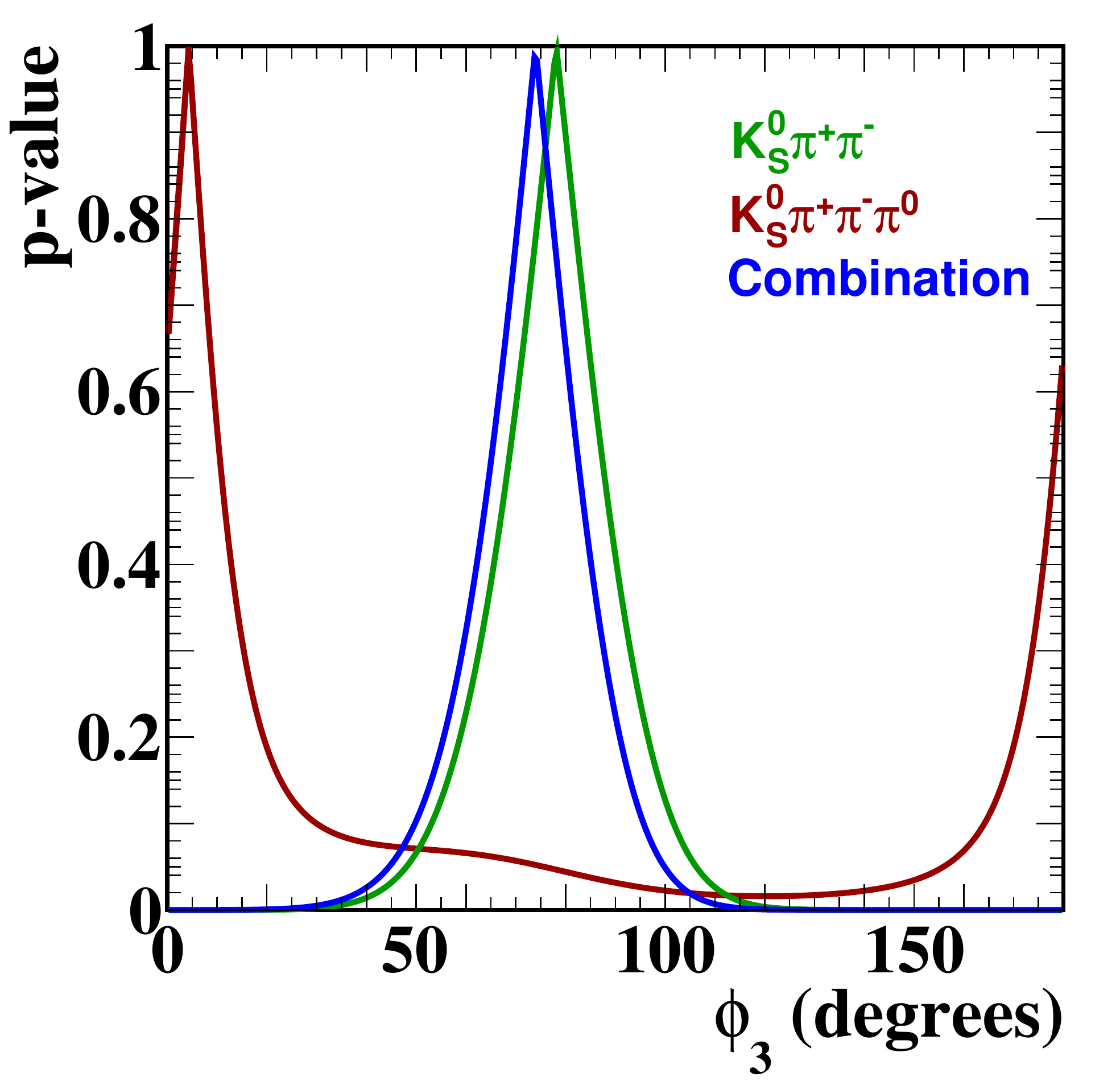}
\caption{Distribution of p-value for $\phi_3$ from multibody $D$ final states at Belle, which is shown by the solid blue curve. The results from $B\to DK^{(*)}$ decays with $D\to K_{\rm S}^0\pi^+\pi^-$ are shown by the solid green curve and the $D\to K_{\rm S}^0\pi^+\pi^-\pi^0$ final states are shown by the solid brown curve~\cite{Belle-GGSZ,Belle-GGSZ2}.}\label{Fig:BelleGGSZ1}
\end{figure}

\begin{figure}[t]
\centering
\includegraphics[width=0.5\columnwidth]{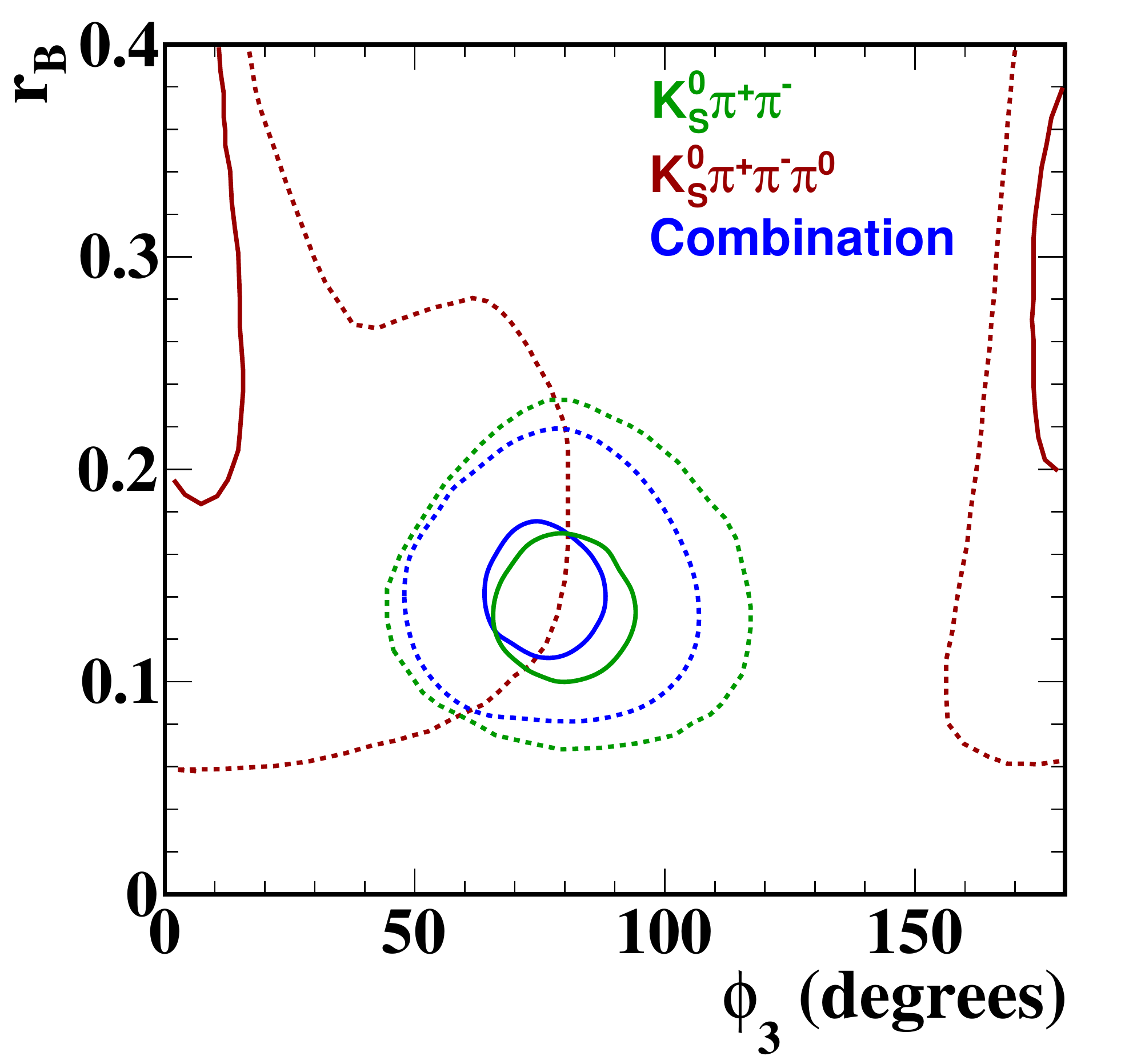}
\caption{Projections of the confidence contours in the $\phi_3 - r_B$ plane from multibody $D$ final states at Belle, which is shown by the blue contours. The results from $B\to DK^{(*)}$ decays with $D\to K_{\rm S}^0\pi^+\pi^-$ are shown by the green contours and the $D\to K_{\rm S}^0\pi^+\pi^-\pi^0$ final states are shown by the brown contours. The solid and dashed curves correspond to one and two standard deviation contours, respectively~\cite{Belle-GGSZ,Belle-GGSZ2}. }\label{Fig:BelleGGSZ2}
\end{figure}

We combine the results presented here with the model-independent $B^{+}\to~D(K_{\rm S}^0\pi^+\pi^-)K^{+}$~\cite{Belle-GGSZ} and $B^0\to~D^0(K_{\rm S}^0 \pi^+ \pi^-) K^{*0}$~\cite{Belle-GGSZ2} results from Belle. Without our measurement, the combination leads to $\phi_3~=~(78^{+14}_{-15})^{\circ}$. Including our measurement, the combination gives $\phi_3~=~(74^{+13}_{-14})^{\circ}$. The distributions of p-values for the $\phi_3$ measurements from the individual $D$ final states and the combination are given in figure~\ref{Fig:BelleGGSZ1}. The separate measurements and the combination likelihood contours in the $(\phi_3, r_B)$ plane are shown in figure~\ref{Fig:BelleGGSZ2}. 

\section{Conclusion}
\label{Sec:con}
We have performed the first measurement of the unitarity triangle angle $\phi_3$ using a model-independent analysis of $B^{+} \to D(K_{\rm S}^0\pi^+\pi^-\pi^0)K^{+}$ decays using  the full data sample collected by the Belle detector, which corresponds to $772\times 10^6$ $B\overline{B}$ events. The $D$ strong-phase difference measurements for $D\to K_{\rm S}^0\pi^+\pi^-\pi^0$ \cite{Resmi} are used as external inputs to the analysis. The result obtained is $\phi_3 = (5.7~^{+10.2}_{-8.8} \pm 3.5 \pm 5.7)^{\circ}$. The first uncertainty is statistical, the second is systematic, and the third is due to the uncertainty on the $c_i$ and $s_i$ measurements. The ratio of the suppressed and favoured amplitudes is $r_{B} = 0.323 \pm 0.147 \pm 0.023 \pm 0.051$. 

 This measurement can be improved upon once a suitable amplitude model for $D^{0}\to K_{\rm S}^0\pi^{+}\pi^{-}\pi^{0}$ is available to provide guidance in choosing a more sensitive binning. Furthermore, the larger sample of $e^{+}e^{-}\to \psi(3770)$ data that has been collected by BESIII will determine $c_i$ and $s_i$ more precisely, thus reducing the systematic uncertainty. The results presented here, combined with the improvements in binning and the increased sample of $B$ decays that will be available at Belle~II, mean that  model-independent analysis of $B^{+} \to D(K_{\rm S}^0\pi^+\pi^-\pi^0)K^{+}$ is a very promising addition to the suite of modes to be used to determine $\phi_3$ to a precision of 1\textendash 2$^{\circ}$ \cite{b2tip}. 

\acknowledgments
We thank the KEKB group for the excellent operation of the
accelerator; the KEK cryogenics group for the efficient
operation of the solenoid; and the KEK computer group, and the Pacific Northwest National
Laboratory (PNNL) Environmental Molecular Sciences Laboratory (EMSL)
computing group for strong computing support; and the National
Institute of Informatics, and Science Information NETwork 5 (SINET5) for
valuable network support.  We acknowledge support from
the Ministry of Education, Culture, Sports, Science, and
Technology (MEXT) of Japan, the Japan Society for the 
Promotion of Science (JSPS), and the Tau-Lepton Physics 
Research Center of Nagoya University; 
the Australian Research Council including grants
DP180102629, 
DP170102389, 
DP170102204, 
DP150103061, 
FT130100303; 
Austrian Science Fund (FWF);
the National Natural Science Foundation of China under Contracts
No.~11435013,  
No.~11475187,  
No.~11521505,  
No.~11575017,  
No.~11675166,  
No.~11705209;  
Key Research Program of Frontier Sciences, Chinese Academy of Sciences (CAS), Grant No.~QYZDJ-SSW-SLH011; 
the  CAS Center for Excellence in Particle Physics (CCEPP); 
the Shanghai Pujiang Program under Grant No.~18PJ1401000;  
the Ministry of Education, Youth and Sports of the Czech
Republic under Contract No.~LTT17020;
the Carl Zeiss Foundation, the Deutsche Forschungsgemeinschaft, the
Excellence Cluster Universe, and the VolkswagenStiftung;
the Department of Science and Technology of India; 
the Istituto Nazionale di Fisica Nucleare of Italy; 
National Research Foundation (NRF) of Korea Grants
No.~2016R1D1A1B01010135,  No.~2016R1D1A1B02012900,  No.~2018R1A2B3003643, 
No.~2018R1A6A1A06024970, \\ No.~2018R1D1A1B07047294,  No.~2019K1A3A7A09033840;
Radiation Science Research Institute, Foreign Large-size Research Facility Application Supporting project, the Global Science Experimental Data Hub Center of the Korea Institute of Science and Technology Information and KREONET/GLORIAD;
the Polish Ministry of Science and Higher Education and 
the National Science Center;
the Grant of the Russian Federation Government, Agreement No.~14.W03.31.0026; 
the Slovenian Research Agency;
Ikerbasque, Basque Foundation for Science, Spain;
the Swiss National Science Foundation; 
the Ministry of Education and the Ministry of Science and Technology of Taiwan;
and the United States Department of Energy and the National Science Foundation.

\appendix
\section{Efficiency and Migration matrix}
\label{app:migmatrix}
The efficiencies in nine bins of $D$ phase space in $D^{*\pm} \to D\pi^{\pm}$, $B^{\pm}\to DK^{\pm}$ and $B^{\pm}\to D\pi^{\pm}$ decays determined from signal MC samples are given in table~\ref{Table:bineff}.
\begin{table} [t] 
\centering 
 \begin{tabular} {c c ccc cc}
\hline 
Bin & \multicolumn{6}{c}{$\epsilon$ (\%)} \\[0.5ex]
\hline
& $D^{*+} \to D^0 \pi^+$ & $D^{*-} \to \overline{D}^0\pi^-$& $B^{+}\to DK^+$ & $B^{-}\to DK^-$ & $B^+\to D\pi^+$ & $B^-\to D\pi^-$\\[0.5ex]
\hline
\hline
1 &  3.07$\pm$0.06 & 3.02$\pm$0.06 & 3.77$\pm$0.05 & 3.84$\pm$0.05 & 4.43$\pm$0.06 & 4.35$\pm$0.06\\[0.5ex]
2 &  3.77$\pm$0.05 & 4.83$\pm$0.09 & 5.44$\pm$0.07 & 5.01$\pm$0.04 & 6.15$\pm$0.08 & 5.47$\pm$0.04\\[0.5ex]
3 &  5.66$\pm$0.14 & 3.66$\pm$0.05 & 4.97$\pm$0.04 & 4.88$\pm$0.10 & 5.55$\pm$0.05 & 5.55$\pm$0.10\\[0.5ex]
4 &  3.60$\pm$0.11 & 3.72$\pm$0.12 & 4.55$\pm$0.10 & 4.63$\pm$0.09 & 5.29$\pm$0.11 & 5.17$\pm$0.09\\[0.5ex]
5 &  3.77$\pm$0.14 & 3.38$\pm$0.11 & 4.89$\pm$0.10 & 4.28$\pm$0.10 & 5.47$\pm$0.10 & 4.53$\pm$0.11\\[0.5ex]
6 &  3.71$\pm$0.11 & 3.45$\pm$0.11 & 4.68$\pm$0.09 & 4.28$\pm$0.09 & 5.46$\pm$0.10 & 5.04$\pm$0.09\\[0.5ex]
7 &  3.87$\pm$0.17 & 4.03$\pm$0.19 & 4.92$\pm$0.16 & 4.66$\pm$0.14 & 5.64$\pm$0.18 & 5.29$\pm$0.14\\[0.5ex]
8 &  3.36$\pm$0.24 & 3.53$\pm$0.21 & 5.36$\pm$0.19 & 4.77$\pm$0.20 & 5.75$\pm$0.20 & 5.56$\pm$0.22\\[0.5ex]
9 &  3.32$\pm$0.16 & 3.21$\pm$0.16 & 4.64$\pm$0.14 & 4.21$\pm$0.13 & 4.87$\pm$0.15 & 4.83$\pm$0.14\\[0.5ex]

\hline
\end{tabular}
\caption{Efficiency in each bin of the $D$ phase space for $D^{*\pm} \to D\pi^{\pm}$, $B^{\pm}\to DK^{\pm}$, and $B^{\pm}\to D\pi^{\pm}$ decays determined from the corresponding signal MC samples.}\label{Table:bineff}
\end{table} 
The migration matrices for $B^+\to DK^+$ and $B^+\to D\pi^+$ decays estimated from signal MC samples are given in tables~\ref{Table:MigDK} and \ref{Table:MigDpi}, respectively.
\begin{table} [t] 
\centering  
 \begin{tabular} {l ccc cccccc }
\hline 
Bin no.& 1 & 2& 3& 4& 5& 6& 7& 8& 9 \\[0.5ex]
\hline
\hline
1 & 0.93 & 0.01 & 0.01 & 0.01 & 0.01 & 0.01 & 0.00 & 0.00 & 0.01\\[0.5ex]
2 & 0.01 & 0.96 & 0.02 & 0.00 & 0.00 & 0.00 & 0.00 & 0.00 & 0.00 \\[0.5ex]
3 & 0.01 & 0.02 & 0.95 & 0.00 & 0.00 & 0.00 & 0.00 & 0.00 & 0.00 \\[0.5ex]
4 & 0.04 & 0.03 & 0.02 & 0.90 & 0.00 & 0.00 & 0.00 & 0.00 & 0.01 \\[0.5ex]
5 & 0.04 & 0.01 & 0.03 & 0.01 & 0.91 & 0.01 & 0.00 & 0.00 & 0.01 \\[0.5ex]
6 & 0.02 & 0.02 & 0.01 & 0.01 & 0.01 & 0.92 & 0.01 & 0.00 & 0.00 \\[0.5ex]
7 & 0.01 & 0.03 & 0.02 & 0.00 & 0.01 & 0.02 & 0.91 & 0.00 & 0.01 \\[0.5ex]
8 & 0.01 & 0.02 & 0.02 & 0.01 & 0.00 & 0.01 & 0.01 & 0.88 & 0.02 \\[0.5ex]
9 & 0.06 & 0.02 & 0.02 & 0.01 & 0.01 & 0.01 & 0.01 & 0.00 & 0.86 \\[0.5ex]
\hline
\end{tabular}
\caption{Migration matrix for $B^+\to DK^+$ decays estimated from the signal MC sample. The rows correspond to the true bins and columns show the reconstructed bins.\label{Table:MigDK}}
\end{table}

\begin{table} [t] 
\centering  
 \begin{tabular} {l ccc cccccc }
\hline 
Bin no.& 1 & 2& 3& 4& 5& 6& 7& 8& 9 \\[0.5ex]
\hline
\hline
1 & 0.93 & 0.01 & 0.01 & 0.01 & 0.01 & 0.01 & 0.00 & 0.00 & 0.01\\[0.5ex]
2 & 0.01 & 0.96 & 0.01 & 0.00 & 0.00 & 0.00 & 0.00 & 0.00 & 0.00 \\[0.5ex]
3 & 0.01 & 0.02 & 0.95 & 0.01 & 0.00 & 0.01 & 0.00 & 0.00 & 0.00 \\[0.5ex]
4 & 0.03 & 0.02 & 0.02 & 0.92 & 0.00 & 0.01 & 0.00 & 0.00 & 0.00 \\[0.5ex]
5 & 0.03 & 0.02 & 0.02 & 0.01 & 0.91 & 0.01 & 0.00 & 0.00 & 0.01 \\[0.5ex]
6 & 0.03 & 0.02 & 0.01 & 0.01 & 0.00 & 0.93 & 0.00 & 0.01 & 0.00 \\[0.5ex]
7 & 0.01 & 0.03 & 0.01 & 0.00 & 0.00 & 0.01 & 0.92 & 0.00 & 0.01 \\[0.5ex]
8 & 0.00 & 0.01 & 0.03 & 0.00 & 0.01 & 0.01 & 0.01 & 0.92 & 0.01 \\[0.5ex]
9 & 0.05 & 0.01 & 0.01 & 0.01 & 0.00 & 0.02 & 0.01 & 0.01 & 0.88 \\[0.5ex]
\hline
\end{tabular}
\caption{Migration matrix for $B^+\to D\pi^+$ decays estimated from the signal MC sample. The rows correspond to the true bins and columns show the reconstructed bins.\label{Table:MigDpi}}
\end{table}

\end{document}